# Exploring acceptance of autonomous vehicle policies using KeyBERT and SNA: Targeting engineering students

Jinwoo Ha[a], Dongsoo Kim[b*]

[a]Department of IT Logistics and Distribution, Soongsil University, 369 Sangdo-ro, Dongjak-gu, Seoul, 06978, Republic of Korea
[b]Department of Industrial and Information Systems Engineering, Soongsil University, 369 Sangdo-ro, Dongjak-gu, Seoul, 06978, Republic of Korea

**Abstract**

This study aims to explore user acceptance of Autonomous Vehicle (AV) policies with improved text-mining methods. Recently, South Korean policymakers have viewed Autonomous Driving Car (ADC) and Autonomous Driving Robot (ADR) as next-generation means of transportation that will reduce the cost of transporting passengers and goods. They support the construction of V2I and V2V communication infrastructures for ADC and recognize that ADR is equivalent to pedestrians to promote its deployment into sidewalks. To fill the gap where end-user acceptance of these policies is not well considered, this study applied two text-mining methods to the comments of graduate students in the fields of "Industrial," "Mechanical," and "Electronics · Electrical · Computer." One is the Co-occurrence Network Analysis (CNA) based on TF-IWF and Dice coefficient, and the other is the Contextual Semantic Network Analysis (C-SNA) based on both KeyBERT, which extracts keywords that contextually represent the comments, and double cosine similarity. The reason for comparing these approaches is to balance interest not only in the implications for the AV policies but also in the need to apply quality text mining to this research domain. Significantly, the limitation of frequency-based text mining, which does not reflect textual context, and the trade-off of adjusting thresholds in Semantic Network Analysis (SNA) were considered. As the results of comparing the two approaches, the C-SNA provided the information necessary to understand users' voices using fewer nodes and features than the CNA. The users who pre-emptively understood the AV policies based on their engineering literacy and the given texts revealed potential risks of the AV accident policies. This study adds suggestions to manage these risks to support the successful deployment of AVs on public roads.

## Declaration of interest

The first author worked for Neubility, an autonomous driving robot manufacturer, from April 26, 2021, to June 11, 2021. During this time, the first author was involved in a project related to applying for Regulatory Exceptions for Demonstration to the Korean government. This project aimed to provide an official demonstration opportunity before Neubility commercialized its product. However, the first author left the company and did not participate in the actual demonstration process. This short work experience was not an inappropriate influence on this study, which was conducted based on systematic approaches and facts.

## 1. Introduction

Autonomous Vehicle (AV) is an emerging technology that enables expectations such as reducing transportation costs, enhancing mobility, and decreasing vehicle collisions. However, if manufacturers bear substantial liability for accidents, the technology adoption may be delayed. Conversely, if manufacturers take weak liability, developing safe products and achieving the societal goal of providing appropriate compensation to victims may fail. Therefore, policymakers need to strike a balance between the two. (Anderson et al., 2016)

Recently, AVs have expanded their forms and functions from passenger to freight and cargo transport. (Jones et al., 2023) The ground types of AVs are classified based on the road infrastructures they use. (Li, J. et al., 2021) One is the

* Corresponding author. (dskim@ssu.ac.kr)

Autonomous Driving Car (ADC), which utilizes carriageways, and the other is the Autonomous Driving Robot (ADR), which utilizes sidewalks. This study is interested in both.

Exploring user concerns is essential for deploying ADC on a large scale on public roads. (Das et al., 2019) ADC, which is difficult for users to accept, can take longer to be recognized as an innovation. (Kohl et al., 2017) User concerns about ADR can also be explored in advance to reference for improvements in ADR design. (Io and Lee, 2019) While South Korean policymakers are proactive in promoting the deployment of AVs on public roads, they are relatively passive in exploring user concerns and attributing appropriate liability to AV-related companies.

In the case of ADC, the South Korean government has a roadmap to commercialize Full Self Driving (FSD) by 2027 through the construction of V2I (Vehicle to Infrastructure) and V2V (Vehicle to Vehicle) communication infrastructures in the country. (Ministry of Land, Infrastructure and Transport, 2022) In this context, Hyundai is testing robotic taxis in Seoul that overcome traffic signal perception limitations of sensors through wireless communication with traffic signal systems and is planning to expand them to North America soon. (Park, K. I., 2022; Kim, C. S., 2023)

Sharing information about destinations and routes among vehicles can alleviate the non-cooperative traffic pattern known as user equilibrium (Bagloee et al., 2016), and integrating ADC into Cooperative Intelligent Transportation Systems (C-ITS) can enable more efficient and safer autonomous driving experiences. (Uhlemann, 2015; Premebida et al., 2018; Naranjo et al., 2020) However, while the potential for manufacturers' advertising to exceed actual levels of their technologies and benefits to users during transitioning to FSD should be considered, it is challenging to find evidence of such consideration by the policymakers.

For example, Tesla has advertised that they can achieve pure vision-based FSD using only cameras and autonomous driving software. However, this company has been criticized for using the term “FSD” in a way inconsistent with the SAE J3016 standard, indicating exaggerated advertising. Moreover, some drivers of Tesla's ADCs have experienced phantom braking. (Kolodny, 2021; Templeton, 2022; Shuttleworth, 2019; McFarland, 2023; Stempel, 2022; Kolodny, 2022; Mitchell, 2023; Gonzalez, 2022)

In this study, the contrasting cases of Hyundai and Tesla were presented to users, and an open-ended question was posed about overall forecasts for the realization of FSD in ADC. (See Appendix A.) What concerns can be observed from the users' responses? This was the first research question (RQ 1) investigated by text mining their comments.

**Table 1.** The change in punishment levels for ADR operators in accidents with pedestrians

| Time Points | Status of ADR | Entering Sidewalk | Act Subjects to Punishment | Punishment Levels |
|---|---|---|---|---|
| Before amendment | Vehicle | Illegal | Causing pedestrian death or injury | imprisonment without prison labor for not more than five years, a fine not exceeding 20 million KRW |
| After amendment | Pedestrian | Legal | Causing danger or impediment to pedestrian | a fine not exceeding 200,000 KRW, misdemeanor imprisonment, or a minor fine |

In the case of ADR, the Korean National Assembly recognized ADR as being almost equal to pedestrians and permitted its entry onto sidewalks. (National Assembly Secretariat, 2023a) This recognition is not a new concept, as some states in the United States, such as Pennsylvania, have already implemented it. (Blanco, 2021) San Francisco is one of the few exceptions that strictly regulates ADR. (Said and Evangelista, 2018) However, it is still necessary to consider potential concerns about ADR. The recognition of ADR as equal to pedestrians holds significant meaning.

Because ADR is not a vehicle in a legal sense and punishment levels imposed on operators in accidents with pedestrians are significantly reduced compared to before. (See Table 1.) In cases of accidents caused by product defects, manufacturers are only obligated to provide compensation for the damages. (Chairperson of the National Assembly Committee on Public Administration and Security, 2023; Office For Government Policy Coordination, 2022) Moreover, the government has indicated its intention to introduce a safety certification system, planned to allow ADR specifications to a maximum weight of 100kg and a maximum speed of 15km/h. (National Assembly Secretariat, 2023b) As the Korean National Assembly entrusted this decision to the government (National Assembly Secretariat, 2023c), it has now become possible for ADRs, faster than pedestrians, to appear on urban sidewalks.

Of course, a minority of policymakers expressed safety concerns. (Park, K. C., 2022; National Assembly Secretariat, 2023b; National Assembly Secretariat, 2023d) However, in Korea, some claims have been spreading through ADR manufacturers and industry-related media, including that the regulations on ADR have been hindering industrial growth (Lee, N. Y., 2021; Namgoong, M. K., 2022), that ADR technologies will inevitably advance as driving data accumulates (Choi, A. R., 2022), and that ADR, unlike scooter, handles short-range delivery and does not need to move quickly, making it safer. (Yoon, H. S., 2021; Kim, C. H., 2022)

Therefore, the concerns of the minority of policymakers did not receive much attention, and the amendment was ultimately implemented. So, how will the users respond to this policy change? This is the second research question (RQ 2), and no study within the perceivable scope has explored this. To fill the gap, the given texts presented contrasting views on the change (See Appendix A.), and an open-ended question was posed to solicit the users' opinions. RQ 2 was investigated by conducting text mining on their comments.

The overall sequence for answering RQs 1 and 2 is as follows: In Literature Review, the trend of previous studies exploring AV user acceptance by text mining and the contributions of this study are presented. In Methods, the Co-occurrence Network Analysis (CNA) and the Contextual Semantic Network Analysis (C-SNA) are presented to be applied in parallel for comparison. Data collection and Preprocessing describe the process of collecting and preprocessing data to apply these methods, and Analysis and Results report the application results. Comparison and Summary show that C-SNA provides necessary information using fewer nodes and features than CNA. Discussions provide suggestions for managing the risks of the AV policies implied by the answers to RQs 1 and 2, and Conclusions present the findings, limitations, and future work of this study.

## 2. Literature review

Text mining, often used to explore user acceptance, also has been consistently used in studies of ADC users across transportation-related research domains. (See Table 2.) However, studies on ADR users and user acceptance of the South Korean AV policy are scarce. This study makes its first contribution by filling these gaps.

**Table 2.** The previous studies on AV user acceptance using text mining

| Authors | Categories | Data sources | Methods |
|---|---|---|---|
| Xing et al., 2023 | ADC | Online Survey on Lucid (752 participants) | Sentiment analysis, co-occurrence analysis, clustering |
| Dos Santos et al., 2022 | ADC | EU-27+UK news articles (19,540 articles), Eurobarometer Survey 496 (27,565 citizens) | Sentiment analysis, clustering, LCA |
| Ding et al., 2021 | ADC | Twitter (696,835 tweets) | Sentiment analysis, LDA topic modeling, time series analysis, user analysis |
| Das, 2021 | ADC | BikePGH online survey (795 participants) | Association rules mining, LDA topic modeling |
| Io and Lee, 2019 | ADR | Weibo (1,814 comments) | Sentiment analysis, LDA topic modeling |
| Das et al., 2019 | ADC | Youtube (25,629 comments) | Sentiment analysis, TF-IDF |
| Li et al., 2018 | ADC | Youtube (50,000 comments) | Sentiment analysis, word cloud |
| Khol et al., 2017 | ADC | Twitter (601,778 tweets) | Classification model, time series analysis |

This study also recognizes the importance of understanding the contexts of users' voices to comprehend their acceptance. So far, previous studies have primarily applied techniques that categorize acceptance of AV users into 2-4 elements, such as sentiment analyses (Xing et al., 2023; Dos Santos et al., 2022; Ding et al., 2021; Io and Lee, 2019; Das et al., 2019; Li et al., 2018) and classification model. (Khol et al., 2017) Therefore, there were advantages in interpreting analysis results intuitively, but there were limitations in understanding the broader contexts of users.

It is true that previous researchers were not satisfied with exploring user acceptability in fewer elements and tended to use additional text-mining methods. The following methods have been employed: CNA to explore correlations among terms (Xing et al., 2023), calculating TF-IDF (Spärck Jones., 1972) to identify rare but important trends (Das et al., 2019), exploring clusters of participants or documents around specific terms (Xing et al., 2023; Dos Santos et al., 2022), times series analyses (Ding et al., 2021; Khol et al., 2017) or comparing word clouds before and after an AV accident (Li et al., 2018) to understand user acceptability in a temporal context, Latent Dirichlet Allocation (Blei et al., 2003) to extract topics from AV users' voices. (Ding et al., 2021; Das, 2021; Io and Lee, 2019)

These various efforts show that the ultimate interest of researchers lies in finding the actual contexts from the users' voices. Especially, LDA (Latent Dirichlet Allocation) has been regarded as a popular unsupervised topic modeling algorithm (Chaney and Blei, 2012; Sridhar, 2015; Suominen and Tovianen, 2016), so it is not surprising that it has been widely utilized in previous studies second most utilized after sentiment analysis. However, LDA shows a limitation of frequency-based text mining methods: ignorance of context.

LDA is not an algorithm that pays attention to the contexts of texts but is based on probabilistic assumptions about the distribution of words and topics. Thus, as Harris (1954) early on likened to putting words into a bag and mixing them up, calling it “a Bag of words” (Harris, 1954), it assumes terms position exchangeability for computational efficiency. (Blei et al., 2003) For example, the difference between the two sentences, “Transportation Research Part A accepted my paper.” and “My paper accepted Transportation Research Part A.” is ignored.

One of the solutions is utilizing a pre-trained Large Language Model (LLM) for contextual embeddings. The well-known research on Transformers with attention mechanisms has been widely cited (Vaswani et al., 2017), and the same goes for research on BERT (Bidirectional Encoder Representations from Transformers), which was pre-trained after separating the encoder from Transformer. (Devlin et al., 2019) Variations of BERT in Korean have also proven the ability of LLM to reflect contexts of texts in embeddings. (Park, J. W., 2019; Lee, J. B., 2020; Lee, S. A. et al., 2020; Lee, J. B., 2021; Lee, H. J. et al., 2021; Park, S. J. et al., 2021; Park, J. W. and Kim, D. G., 2021)

BERTopic is a utilization of contextual embeddings of BERT models and has demonstrated superior performance to LDA in the evaluation of topic coherence (Bouma, 2009; Lau et al., 2014), topic diversity (Dieng et al., 2014) on 20 Newsgroups dataset. (Grootendorst, 2022a) However, this topic modeling framework includes the dimension reduction algorithm UMAP (McInnes et al., 2018), which remains very few topics for data with less than 1,000 entries. (Grootendorst, 2022b) In the AV acceptance research domain, collecting 1,000 data entries is relatively feasible from social media (Ding et al., 2021; Io and Lee., 2019; Das et al., 2019, Li et al., 2018; Khol et al., 2017), but with few exceptions (Dos Santos et al., 2022), it is not accessible from surveys. (Xing et al., 2023; Das, 2021)

Fortunately, KeyBERT, developed by Grootendorst (2020), can extract keywords based on context embeddings, even with a single document. (Grootendorst, 2020) In this research domain, interviews are often conducted with dozens of participants. (Sindi and Woodman, 2021; Martinho et al., 2021; Hilgarter and Granig, 2020; Merfeld et al., 2019) Therefore, promoting the utilization of KeyBERT is expected to support quantifying qualitative data.

This study has shown that it is possible to extract Contextual Semantic Networks based on the context-condensed keywords extracted from the users' comments using KeyBERT. This approach is the second contribution of this study and not only pays attention to the actual contexts of the users but also proves efficient in interpreting information. Contextual Semantic Networks provided the necessary information to understand the contexts of the participants' comments with fewer nodes and features than Co-occurrence Networks.

## 3. Methods

The process of this study is shown in Fig 1. First, we collected data through an email survey. (Step 1) The end user population of this study was defined as graduate students in the fields of “Industry,” “Mechanical,” and “Electronics·Electrical·Computer” from 26 universities located in Seoul, South Korea. (See 4.1) Survey participants understood the issues presented in the Introduction through the given texts in the questionnaire (See Appendix A.) and submitted their opinions as open-ended responses. The comments received from the users, who pre-emptively understood the AV policies based on their engineering literacy and the given texts, were expected to suggest potential risks of the policies. We conducted preprocessing suitable for Korean characteristics on the corpus obtained through the survey (Step 2), and the detailed methods are presented in 4.2.

On the other hand, Semantic Network Analysis (SNA) is a widely used method to intuitively understand contexts of texts by representing relationships between keywords as a network. This study explored the users' comments through two SNA approaches. One is Co-occurrence Network Analysis (CNA) to explore the users' superficial voices (Step 3), and the other is Contextual Semantic Network Analysis (C-SNA), which compressively presents the core contexts of their voices. (Step 4b) KeyBERT contributed to extracting the keywords, which are the features used in C-SNA. (Step 4a)

The reason for introducing different SNA approaches to this study is to demonstrate the distinctiveness of C-SNA based on KeyBERT through comparison. Generally, SNA sets thresholds to control the amount of information represented in a network. In this case, there could be a trade-off where it includes much information but is hard to interpret, or it is easy to interpret but does not provide any significant information. Therefore, the key is to compressively reflect only the potentially essential features in a semantic network.

Xing et al. (2023) utilized CNA as that solution (Xing et al., 2023). However, our study demonstrated that it was possible to extract semantic networks that were simpler and contextually representative of users' voices without losing crucial information, even when only the keywords extracted by KeyBERT were used in C-SNA, compared to CNA. The information obtained from these improved networks was summarized to answer RQ 1 and RQ 2, and suggestions were added to support the successful public road deployment of AVs in response to these answers. (Step 5) The main methodological ideas utilized in executing this process are explained in the following sections.

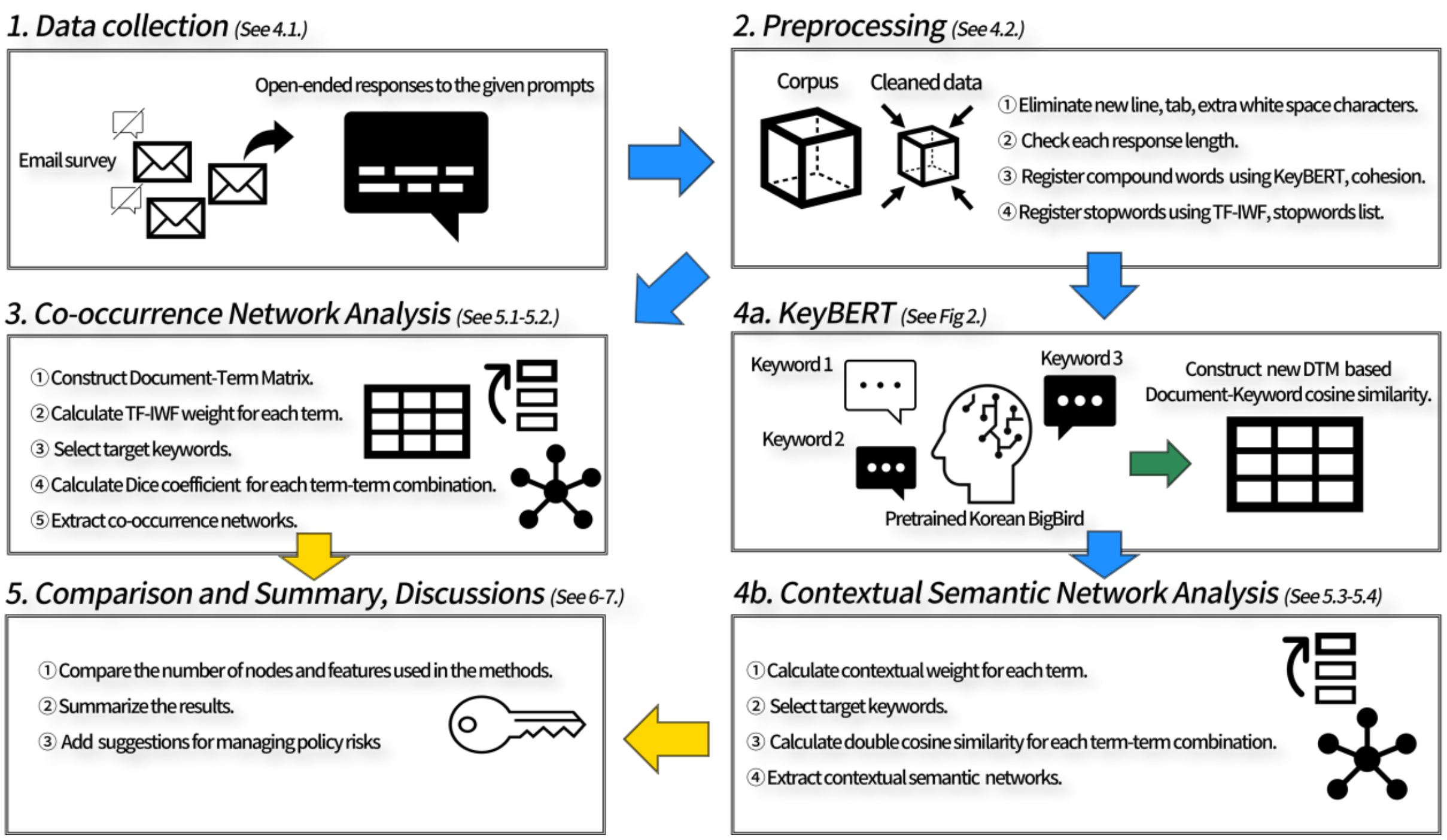

**Fig. 1.** Overall Research Process

### *3.1. Co-occurrence Network Analysis*

Term scoring is a widely used method for selecting keywords for text mining. Xing et al. (2023) used the TF-IDF (Term Frequency-Inverse Document Frequency), proposed by Spärck Jones (1972), to select target keywords that would become the central nodes in co-occurrence networks in their study on AV acceptance. (Spärck Jones, 1972; Xing et al., 2023) Our study inherits their approach in CNA but with a decisive distinction.

This study accepted the proposal to use TF-IWF (Term Frequency-Inverse Word Frequency) instead of TF-IDF for selecting keywords. (Wang et al., 2013; Tian and Wu, 2018; Lu and Jin, 2022) The TF part $tf_{i,j}$ is the frequency $n_{i,j}$ of the $i$-th term $t_i$ appearing in the $j$-th document $d_j$ divided by $\sum_k n_{k,j}$ of all terms appearing in $d_j$. (See Equation 1) Unlike TF-IDF, which takes only $n_{i,j}$ for The TF part, this value reflects the contribution of each term in each document.

$$tf_{i,j} = \frac{n_{i,j}}{\sum_k n_{k,j}} \quad \cdots \quad (1)$$

$$iwf_i = log\frac{\sum_{i=1}^{m} nt_i}{nt_i} \quad \cdots \ (2)$$

$$TF - IWF = \frac{n_{i,j}}{\sum_k n_{k,j}} \times log\frac{\sum_{i=1}^{m} nt_i}{nt_i + 1} \quad \cdots (3)$$

On the other hand, The IWF part, $iwf_i$, is defined as the logarithm of the total frequency $\sum_{i=1}^{m} nt_i$ of all terms appearing in the documents set divided by the total frequency $nt_i$ of the term $t_i$ appearing in the documents set. (See Equation 2.) The value of the IWF part decreases for commonly used terms with high $nt_i$, and it increases for terms that are mentioned in specific document groups with low $nt_i$. Therefore, increasing the influence of terms representing specific document groups is possible while weakening the influence of commonly used terms.

TF-IWF is defined as the product of the TF part and the IWF part (See Equation 3.), which, unlike TF-IDF, can reflect the contribution of each term within each document while weakening the influence of commonly used terms as TF-IDF does. This study used terms with high TF-IWF weights as references for selecting target keywords. The Dice coefficient was utilized for selecting terms around the target keywords. This metric was developed initially by ecologist Dice (1945) to measure inter-species correlations (Dice, 1945) but was found by Rychlý (2008) to be also excellent for exploring term collocations. (Rychlý, 2008) If the number of occurrence cases terms of $x$, $y$ appearing in a corpus are $f_x, f_y$ respectively, and the number of their co-occurrence case is $f_{xy}$, Dice coefficient is as follows:

$$Dice(f_x, f_y) = \frac{2f_{xy}}{f_x + f_y} \quad \cdots (4)$$

Xing et al. (2023) utilized the idea that a collocation consists of base and dependent terms. (Kolesnikova, 2016; Xing et al., 2023) Our study inherits this idea but extracts co-occurrence networks where the weights of the target keywords and surrounding terms nodes are set by TF-IWF, and the edge weights are set by the Dice coefficient.

### *3.2. Cosine similarity calculation based on KeyBERT*

CNA is useful for exploring the superficial voices of users by frequently appearing collocations in their comments. However, like LDA, it assumes terms position exchangeability, so the collocations identified through exploration do not necessarily represent the comments contextually. Therefore, to access the users' core voices, this study used the KeyBERT framework (Grootendorst, 2020) for extracting keywords from the $i$-th document $d_i$.

In Fig 2, $d_i$ is processed in bifurcated flows. In the first flow, $d_i$ is input into a pre-trained Korean BERT model to return contextual embeddings at the document level. In the second flow, Mecab-ko, a Korean morpheme analyzer (Lee, Y. W. and Yoo Y. H., 2013), is specified as the tokenizer parameter for Scikit-learn's CountVectorizer (Pedregosa et al., 2013) to tokenize $d_i$. The candidate tokens that have the potential to represent the document contextually are input into the pre-trained Korean BERT model to return contextual embeddings at the token level.

The two flows converge by measuring the similarity between the document embedding vector and each token embedding vector. The cosine similarity, which is widely used to calculate the similarity between two vectors $A$ and $B$ was used for this measurement:

$$similarity = cos(\theta) = \frac{A \cdot B}{\|A\| \, \|B\|} = \frac{\sum_{i=1}^{n} A_i \times B_i}{\sqrt{\sum_{i=1}^{n}(A_i)^2} \times \sqrt{\sum_{i=1}^{n}(B_i)^2}} \quad \cdots \ (5)$$

KeyBERT filters and remains only the $k_i$ keywords (by default, this is less than or equal to 5) with the highest cosine similarity with each document.

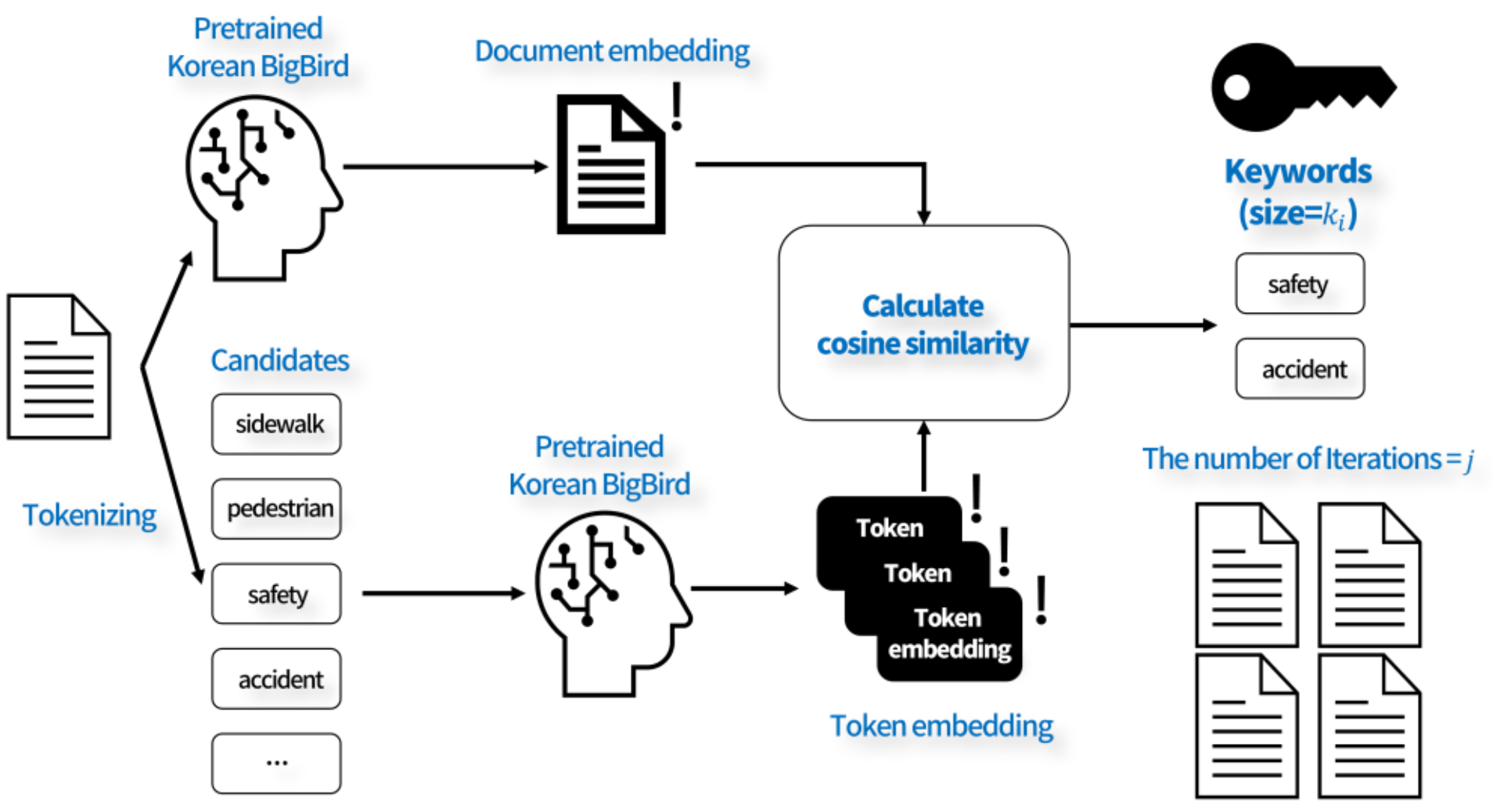

**Fig. 2.** The Process of KeyBERT

Using this process, it is possible to extract keywords that contextually represent each comment. However, considering the space limitations and the nature of this journal, detailed explanations of how contextual embeddings occur within BERT models are replaced with references. As some researchers already provide detailed explanations in the domain of Natural Language Processing, our study avoids tautology. Refer sequentially to the research by Vaswani et al. (2017) on the Transformer (Vaswani et al., 2017), which is based on the Attention mechanism, the core concept of contextual embedding, and the research by Devlin et al. (2019) on BERT, which is a pre-trained model of the encoder independent from the Transformer. (Devlin et al., 2019)

This study adopted KoBigBird-BERT-base as the role of the pre-trained Korean BERT for KeyBERT. (Park, J. W. and Kim, D. G., 2021) BigBird is the model that overcomes BERT's token input length limitation from 512 tokens to 4,096 tokens (Zaheer et al., 2020), and KoBigBird is the pre-trained Korean BigBird on a large corpus collected from general sources such as the National Institute of Korean Language's ModuCorPus (National Institute of Korean Language, 2020), Korean Wikipedia, common crawl, and news data. (Park, J. W. and Kim, D. G., 2021) Therefore, it was expected to return appropriate contextual embeddings for the comments.

### *3.3. Contextual Semantic Network Analysis*

KeyBERT is the method for calculating the similarity between each document and its keywords one by one, so the comprehensive meaning of all comments can be determined by synthesizing the individually extracted results. First, the contextual weight for each keyword was calculated. KeyBERT was iterated for all $j$ comments (See Fig. 2.), and by summing the similarity values of each keyword across all comments, the contextual weight for each keyword in the whole comments was obtained. The weight of the $m$-th keyword out of a total of n keywords is defined as follows:

$$Weight(Keyword_m) = \sum_{i=1}^{j} similarity_{Keyword_m} d_i \quad \cdots \quad (6)$$

Second, the contextual semantic networks were extracted utilizing double cosine similarity. The correlations between each comment and each keyword extracted by KeyBERT were organized as Document-Term Matrices, and the values in the matrix space were the cosine similarity between each comment and each keyword. The new cosine similarity values between each column vector corresponding to each keyword were organized as Term-Term Matrices. The new values in these new matrices represent the correlations between each keyword.

These new values were derived through the inner products that reference the direction in which two different keywords have cosine similarity with the comments. The higher the cosine similarity between the two keywords, the more they contextually represent a specific response group together. The new cosine similarity values between the keywords were distributed in a lower range than the Dice coefficient values because the filtering of KeyBERT presumes competition between terms, and the surviving keywords already condensed their context information.

However, since documents can contain more than one contextual meaning, keywords can have some directional similarity. The contextual semantic networks of this study reflected such similarity. First, target keywords were selected referencing the contextual weights, and then separate semantic networks were created for each of these target keywords. With each target keyword serving as the central node, its surrounding terms served as the surrounding nodes. The weights of each target keyword and its surrounding term nodes are set by contextual weights, and the edge weights are set by the double cosine similarity values.

## 4. Data collection and Preprocessing

This section reports on the data collection and preprocessing performed to apply the methods presented in the previous section.

### *4.1. Data collection*

The data were collected from 15 to 22 March 2023 utilizing the Higher Education in Korea (academyinfo.go.kr), where information on universities in Korea is published. First, the number of graduate students (master's, doctoral, and MS-Ph.D. integrated program students) was counted at each university by the department and registered sex. The registered sex is usually the sex reported to the government at the time of a person's birth, but it can be changed to a different sex if the person applies for a registered sex correction in court.

The departments were clustered into “Industrial,” “Mechanical,” and “Electronics·Electrical·Computer” fields according to the HEK classification standard, identifying the population into six strata by registered sex and field. Next, the emails of 6,891 out of 10,246 members of the population (7,895 males, 2,369 females) were collected by visiting the lab websites for each field and recording the emails in a Google spreadsheet sampling frame. 3,355 empty cells were coded as non-responses. A questionnaire was sent to the collected emails (See Appendix A.), and 407 responses were returned. The participants' field of affiliation was recorded by the researcher's observation, and registered sex was recorded by the participants' self-report. Among the 407 responses, excluding unreliable responses, blanks, and responses where the email address recorded did not match the sampling frame, there were 389 valid responses (average response rate 5.65%) for ADC and 386 valid responses (average response rate 5.60%) for ADR.

**Table 3.** Characteristics of samples for each analysis case

| | ADC (Non-stratified) | ADC (stratified) | ADR (Non-stratified) | ADR (stratified) |
|---|---|---|---|---|
| Responses used | 389 | 283 | 386 | 283 |
| Age-mean | 28.0 | 28.0 | 28.0 | 28.0 |
| Age-std | 4.0 | 4.1 | 4.0 | 4.1 |
| Male | 82.0% | 77.0% | 81.9% | 77.0% |
| Female | 18.0% | 23.0% | 18.1% | 23.0% |
| Industrial | 11.6% | 9.2% | 11.7% | 9.2% |
| Mechanic | 23.4% | 22.3% | 23.3% | 22.3% |
| Electronic, Electrical, Computer | 65.0% | 68.6% | 65.0% | 68.6% |

The valid responses were allocated according to the proportional allocation, commonly used in stratified sampling, where the number of responses assigned to each stratum is proportional to its population size. Allocation was done by generating random numbers and their ranks for each cell in each stratum's sampling frame, including responses starting from the highest-ranked ones until the allocation quota was filled. Because response rates of some strata were lower than average, fewer than valid responses were used in analyses that accurately proportionally matched the size of each stratum to its population size. (See Table 3.)

Exploring as many users' voices as possible is as important as accurately representing the population, so the cases for analysis were classified into four categories depending on the topic and whether they were stratified. Non-stratified cases may have reflected a little more male opinion or fewer opinions from the "Electronics · Electrical · Computer" field, but whether stratified or not did not significantly affect the data analysis results. (See 5.)

### *4.2. Preprocessing*

#### *4.2.1. Length*

Each participant's response in the survey was typically a simple comment of about three lines, so there was no need to divide them into paragraphs or sentences. However, since long texts exceeding KoBigBird's length limit of 4,096 tokens do not get adequately analyzed (Park, J. W. and Kim, D. G., 2021), newline characters, tab characters, and extra spaces were removed, and the length of all comments was tested. Although the character count of 4,096, a stricter standard than tokens, was used as the threshold, no comments exceeded it.

#### *4.2.2. Compound words*

The basic semantic unit in English is a word. So, tokenization can be performed based on spaces using NLTK. (Bird et al., 2009) However, words can be divided into smaller units called morphemes in Korean, so tokenization should not be based on spaces. Typically, a morpheme analyzer suitable for Korean is used as a tokenizer. (Yoo, W. J., 2022) This study used Mecab-ko, which has shown good performance in Korean. (Lee, Y. W. and Yoo, Y. H., 2013; Park, E. L. and Cho, S. J., 2014; Park, K. B. et al., 2020) Mecab-ko originated from the algorithm of Mecab, a morpheme analyzer designed for Japanese, a language that, like Korean, has ambiguous boundaries of semantic units. (Kudo et al., 2004; Kudo, 2006) In this study, a functional limitation of Mecab-ko was compensated with preprocessing.

In Korean, there are free morphemes that can stand alone without combining with other morphemes, such as "자율 autonomous" and "주행 driving." Sometimes these morphemes combine to form compound words, like "자율주행 autonomous driving." The problem is that morpheme analyzers often divide compound words into these free morphemes. The solution is registering each compound word in the user dictionary, prompting Mecab-ko to recognize it as a single token. In this study, an objective algorithm was established to register character strings with high probabilities of being compound words and contextually related to the data in the user dictionary as compound words.

**Algorithm 1.** Main steps of registering compound words into the user dictionary

1. Tokenize each of the $j$ documents into nouns using a morpheme analyzer.
2. Bind the nouns into bigrams and trigrams using the "ngram_range" parameter of CountVectorizer in Scikit-learn.
3. Calculate the contextual weight for each n-gram in the corpus. (See 3.3.)
   3.1. Sort the results in descending order from the n-gram with the highest weight.
4. Starting from the top n-gram, calculate cohesions (See Eq 8.) in the corpus by combining self-constituents.
   4.1. Combinations not appearing in the corpus and returning "KeyError" are excluded.
5. Compare the remaining combinations and register the combination with the highest cohesion.
   5.1. If a combination is already registered, exclude it, and consider the following ranked combination.
   5.2. If all combinations are already registered, do not register any combination.
   5.3. Do not register any combination if there is no superior combination for any combination, including the ones already registered.
6. Iteratively execute steps 4-5 for the next n-grams.
   6.1. The number of iterations is a hyperparameter determined by the researcher depending on the characteristics of the data.

Algorithm 1 can also be explained through an example. The "자율 주행 로봇 autonomous driving robot" in Fig 3 is the n-gram with the highest weight extracted by KeyBERT from 389 ADR comments without noise cleaning in the data. The constituents of this n-gram can be combined in multiple ways. (See Table 4.)

Algorithm 1 merely generalizes the process of identifying which combination should be combined into a compound word based on cohesion and registering them in the user dictionary, such as in Table 4. Cohesion is a cumulative product value of conditional probabilities of the following letters' occurrence as letters are added from the first letter to the right one by one. Strings with high cohesion values are known to be likely to appear as single words in the corpus. (Kim, H. J., 2019) If the conditional probability of letter $c_2$ following $c_1$ is Equation 7, the cohesion of the string $c_1, c_2, \cdots c_{n-1}, c_n$, is defined by Equation 8.

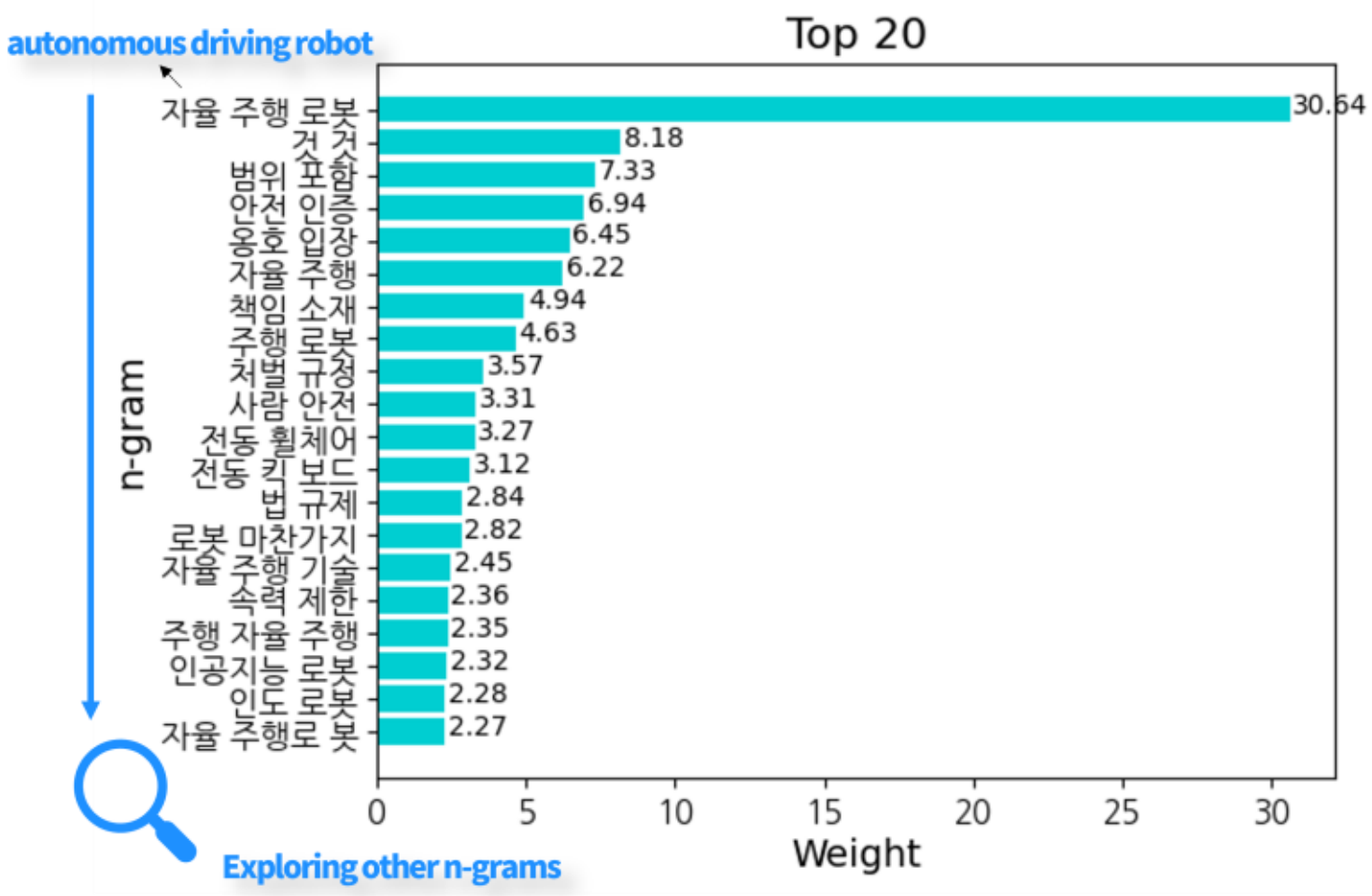

**Fig. 3.** Exploring candidates of compound words

**Table 4.** An example of cohesion calculation

| Combinations | Cohesion | Descriptions |
|---|---|---|
| 자율 autonomous | 0.561 | It is a free morpheme and is already registered. |
| 자율주행 autonomous driving | 0.791 | This is the compound word. |
| 자율주행로봇 autonomous driving robot | 0.487 | The users more often separate "자율주행 autonomous driving" and "로봇 robot" rather than combining them. |
| 주행 driving | 0.670 | It is a free morpheme and is already registered. |
| 주행로봇 driving robot | KeyError | A KeyError is returned because the users do not use it. |
| 로봇 robot | 0.997 | It is a free morpheme and is already registered. |

$$P(c_{1:2}|c_1) = \frac{the\ \ number\ \ of\ \ c_{1:2}}{the\ \ number\ \ of\ \ c_1} \quad \cdots \quad (7)$$

$$cohesion(c_{1:n}) = \left(\prod_{i=1}^{n-1} P(c_{1:i+1}|c_{1:i})\right)^{n-1} \quad \cdots \quad (8)$$

In the calculation of cohesion values in Table 4, SOYNLP, a Python library for Korean preprocessing based on unsupervised learning, was used. (Kim, H. J., 2019) As a result, in the example, "자율주행 autonomous driving" was registered as a compound word. In this study, Algorithm 1 was iterated up to the top 50 n-grams for each analysis case, and 11 compound words were registered in the user dictionary. Thus, KeyBERT can be utilized not only for keyword extraction but also for preprocessing languages with ambiguous word boundaries.

*4.2.3. Stopwords and Affixing Languages*

A corpus typically contains less important terms from an analyst's perspective. These stopwords can be excluded from an analysis using various statistical indicators or a stopwords list. (Vijayarani et al., 2015) In CNA, only terms with high TF-IWF weights or Dice coefficients were used for the analysis, and in C-SNA, stopwords were automatically excluded from the analysis as KeyBERT retains only keywords.

Of course, there were exceptions. Some terms have small IWF values but significant TF values, thus maintaining high TF-IWF values, or were terms that naturally represent the context of comments due to the nature of the research domain. One such example was “자율주행 autonomous driving,” which was registered as a compound word. These terms were compiled into the stopwords list based on a rule. The rule is to compile terms for which the respondents' opinions “about” the terms were important, but not the terms themselves. These terms include “자율주행 autonomous driving,” “완전자율주행 full self-driving,” “완전자율 full self,” “자율 autonomous,” “주행 driving,” “차량 vehicle,” “자동차 car,” “로봇 robot,” “자율주행로봇 autonomous driving robot.”

Next, Korean is an affixing language with many bound morphemes that combine with affixes or postpositions. Verbs and adjectives that are predicates in sentences change their expressions depending on affixes or postpositions. When tokenized, the roots often remain only one letter, making it difficult to infer their meanings. However, nouns and numerals that serve as substantives of sentences are unchanging and free morphemes. So, no matter what is added as an affix or postposition, the root is clear, and the expression is relatively robust. So, in Korean text mining, extracting core meanings from a corpus focusing on nouns is a common approach (Chang, T. Y. et al., 2022; Park, S. T. and Liu, C., 2022), and Korean morpheme analyzers also support functions of extracting only nouns. (Lee, D. G et al., 2010; Lee, Y. W. and Yoo, Y. H., 2013; Ryu, W. H., 2014; Shin, J. S. et al., 2016)

This approach does not excessively increase the abstractness of text-mining analysis results. This is because many nouns in Korean are used like verbs or adjectives with adds of affixes or postpositions; thus, a significant quantity of predicates of sentences are extracted even when only nouns are extracted. This study used general nouns, proper nouns, numerals, and general adverbs that are free morphemes of two letters or more for the analysis. The translations were done to translate Korean nouns into English nouns as much as possible. However, the languages may have different expressions when translated into words or phrases closer to their meanings in English.

## 5. Analysis and Results

CNA and C-SNA were performed in parallel for the preprocessed data. TF-IWF Weights for CNA (See 5.1.) and Contextual Weights for C-SNA (See 5.3.) were calculated. The target keywords were selected based on these weights. Next, Co-occurrence Networks (See 5.2.) and Context Semantic Networks (See 5.4.) were extracted, with the target keywords for each CNA and C-SNA being taken as the central nodes. Since each node in the networks has up to 4-5 edges of high correlations, horizontal line graphs were utilized supplementary for the target keywords to reference the ten neighboring nodes with high correlations. This section provides relatively detailed analysis results, but a summary can be found in the Comparison and Summary. (See Table 8 in 6.)

### *5.1. TF-IWF Weights*

The keywords listed in the top 20 TF-IWF weights are presented in Table 5. When the topic is identical, the keywords appearing in Table 5 seem similar regardless of stratification. This indicates that even non-stratified cases can represent the users' voices.

Despite weakening the influence of common words through TF-IWF, “think” is ranked first in all analysis cases, suggesting that the participants actively expressed their opinions to the extent that “think” was mentioned multiple times in each comment.

In the case of ADC, “realization” and “(im)possible” ranked high, following “think.” This seems to indicate that the respondents were responding to the ask to share their perspectives on the feasibility of FSD of ADC. (See Appendix A.) Although "(im)possible" is a translation of the Korean term “가능” with reserve, it was translated as “possible” in C-SNA using KeyBERT. (See 5.3.) This difference is due to the characteristics of Korean, and detailed reasons are presented in Appendix B. The keyword alternately ahead and behind ranked with “(im)possible” is “tech,” suggesting that the respondents commented on the issue considering the tech level of ADC.

Meanwhile, for the comments on the entry of ADR into sidewalks and mitigation of criminal responsibility of ADR operators, “sidewalk” and “accident” ranked high. It seems that respondents evaluated the issue considering the interaction between ADR, the environment in which it operates, and the problem of accidents. Considering these results, “realization” and “tech” were chosen as target keywords for ADC, and “sidewalk” and “accident” were chosen for ADR. "think" has the highest TF-IWF weight, but as it is a common word and "(im)possible" was still ambiguous, they were excluded from the selection.

**Table 5.** The Top 20 Keywords Ranked by TF-IWF Weights in Each Analysis Case

| Rank | ADC (Non-stratified) | | ADC (Stratified) | | ADR (Non-stratified) | | ADR (Stratified) | |
|---|---|---|---|---|---|---|---|---|
| | Terms | TF-IWF Weights | Terms | TF-IWF Weights | Terms | TF-IWF Weights | Terms | TF-IWF Weights |
| 1 | think | 65.3 | think | 46.6 | think | 55.6 | think | 42.1 |
| 2 | **realization** | **53.2** | **realization** | **37.8** | **sidewalk** | **44.3** | **sidewalk** | **32.8** |
| 3 | (im)possible | 45.4 | **tech** | **32.3** | **accident** | **37.4** | **accident** | **29.5** |
| 4 | **tech** | **43.9** | (im)possible | 31.9 | people | 31.4 | people | 22.0 |
| 5 | problem | 30.3 | problem | 23.9 | pedestrian | 28.4 | pedestrian | 20.9 |
| 6 | advance | 23.1 | accident | 17.0 | tech | 27.6 | tech | 19.7 |
| 7 | accident | 23.1 | advance | 16.6 | safety | 24.4 | safety | 18.3 |
| 8 | road | 19.2 | infra | 15.1 | problem | 24.1 | problem | 18.3 |
| 9 | infra | 19.2 | current | 14.5 | speed | 22.1 | road | 17.1 |
| 10 | situation | 19.0 | road | 14.3 | road | 21.6 | occurrence | 16.0 |
| 11 | current | 18.6 | situation | 13.7 | bicycle | 19.8 | bicycle | 15.7 |
| 12 | sensor | 18.5 | sensor | 13.7 | occurrence | 19.3 | speed | 15.7 |
| 13 | need | 18.1 | time | 13.2 | (im)possible | 18.3 | advance | 13.1 |
| 14 | time | 17.6 | future | 13.0 | advance | 16.7 | (im)possible | 12.1 |
| 15 | future | 16.8 | need | 13.0 | need | 16.7 | need | 11.0 |
| 16 | AI | 14.0 | yet | 10.1 | delivery | 14.7 | delivery | 10.3 |
| 17 | yet | 13.9 | people | 10.0 | support | 14.7 | danger | 10.2 |
| 18 | people | 13.6 | AI | 9.8 | km | 14.3 | km | 10.1 |
| 19 | enough | 13.0 | enough | 9.6 | case | 13.5 | punishment | 9.6 |
| 20 | case | 12.9 | commercial | 9.2 | situation | 13.4 | situation | 9.6 |

### *5.2. Co-occurrence Networks*

To further explore user acceptance of AV, it was investigated what terms each target keyword correlates with. The terms with the top 10 Dice coefficients (DC) with each target keyword are presented in horizontal line graphs (See Fig 4, 6.), and more terms and edges are shown in co-occurrence networks. (See Fig 5, 7.)

Each network includes terms with DC within the top 1.98~2% with each target keyword, and their sizes reflect the TF-IWF weights. Each node has edges up to the top 4 DC terms that range from 0 to 1, and their thicknesses reflect the DC. Some edges that reflect strong correlations within the top 20% of DC in the corpus, including outside each network, were highlighted in red.

Incidentally, each target keyword was highlighted in red, and the colors of neighboring nodes were determined through the widely used Leiden Algorithm, which detects communities among nodes with high connectivity. (Traag et al., 2019) The layouts of networks were determined using the Kamada-Kawai Algorithm, commonly used to optimize graph-theoretical distances between nodes. (Kamada and Kawai, 1989)

#### *5.2.1. Co-occurrence Networks in ADC Cases*

In the CNA results for ADC, as seen in Fig 4, both "realization" and "tech" are included in the top 10 DC terms of each other, regardless of stratification. ($> 0.61$) This suggests that the respondents generally associated FSD realization and tech. Especially in Fig 5, since "tech" and "advance" are thickly connected, it seems that the respondents considered the level of advancement of ADC tech as a precondition for FSD realization. However, in Fig 4, both “realization” and “tech” also have relatively high coefficients with “problem” (> 0.46), and as “problem” is strongly connected with “solve” in Fig 5, it seems that the recognition that there are problems to solve for ADC's transition to FSD was also considerably reflected.

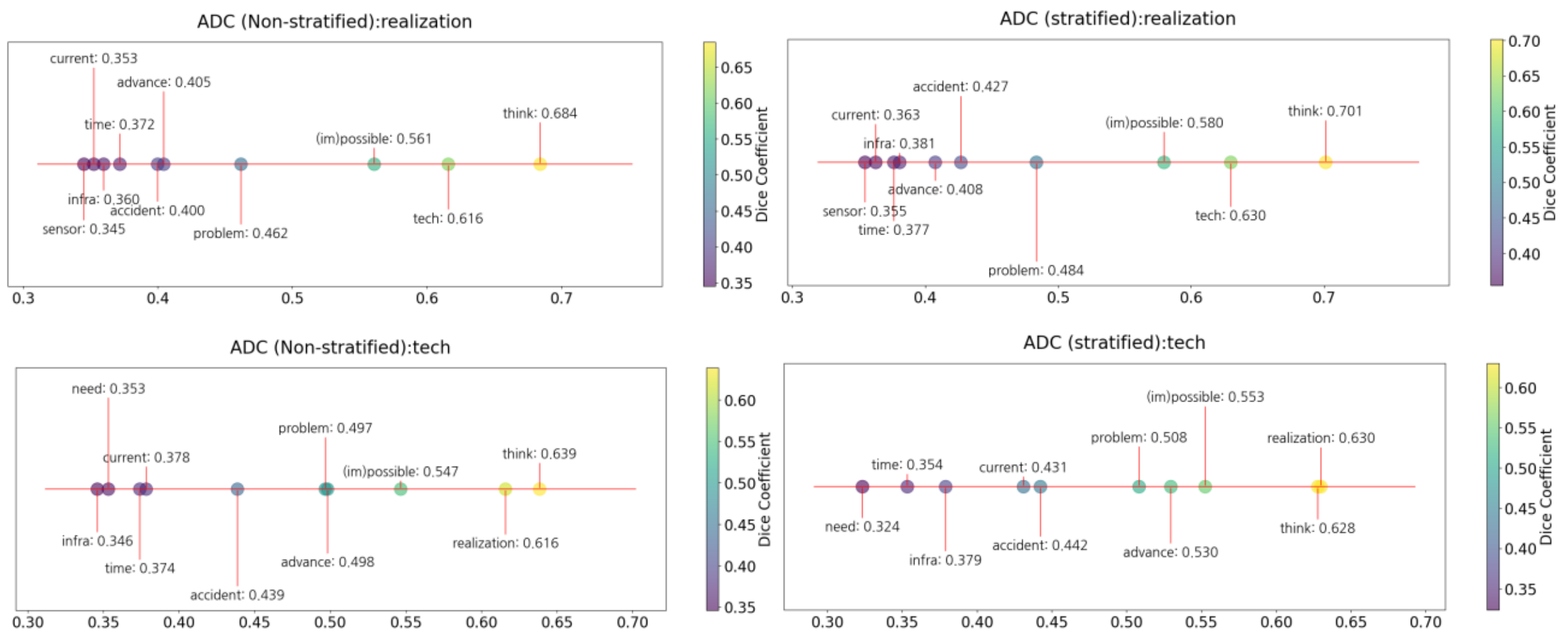

**Fig. 4.** The Top 10 neighboring Terms Ranked by Dice Coefficients in each ADC Case

In Fig 5, “problem” is also connected with “accident.” This connection is not strong enough to be in the top 20%, but it needs to remember that each has edges up to the top 4 Dice coefficients. Therefore, some respondents seem to have primarily associated solving ADC accident problems as a precondition for transitioning to FSD. Moreover, “accident” is strongly connected with “responsibility” in Fig 5. This observation is interesting because it suggests that some engineering students considered social aspects, like responsibility in events of ADC accidents, not just purely technical factors.

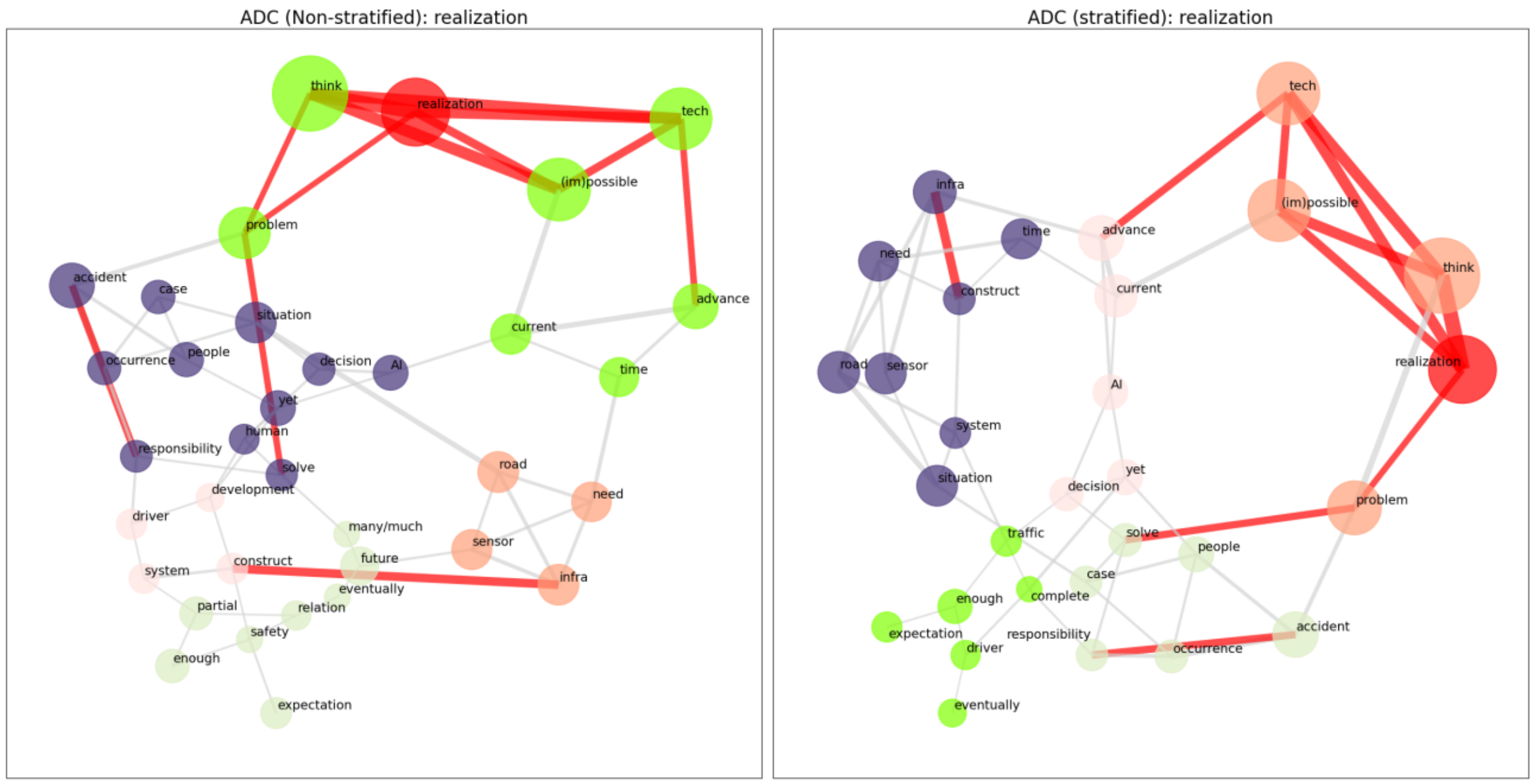

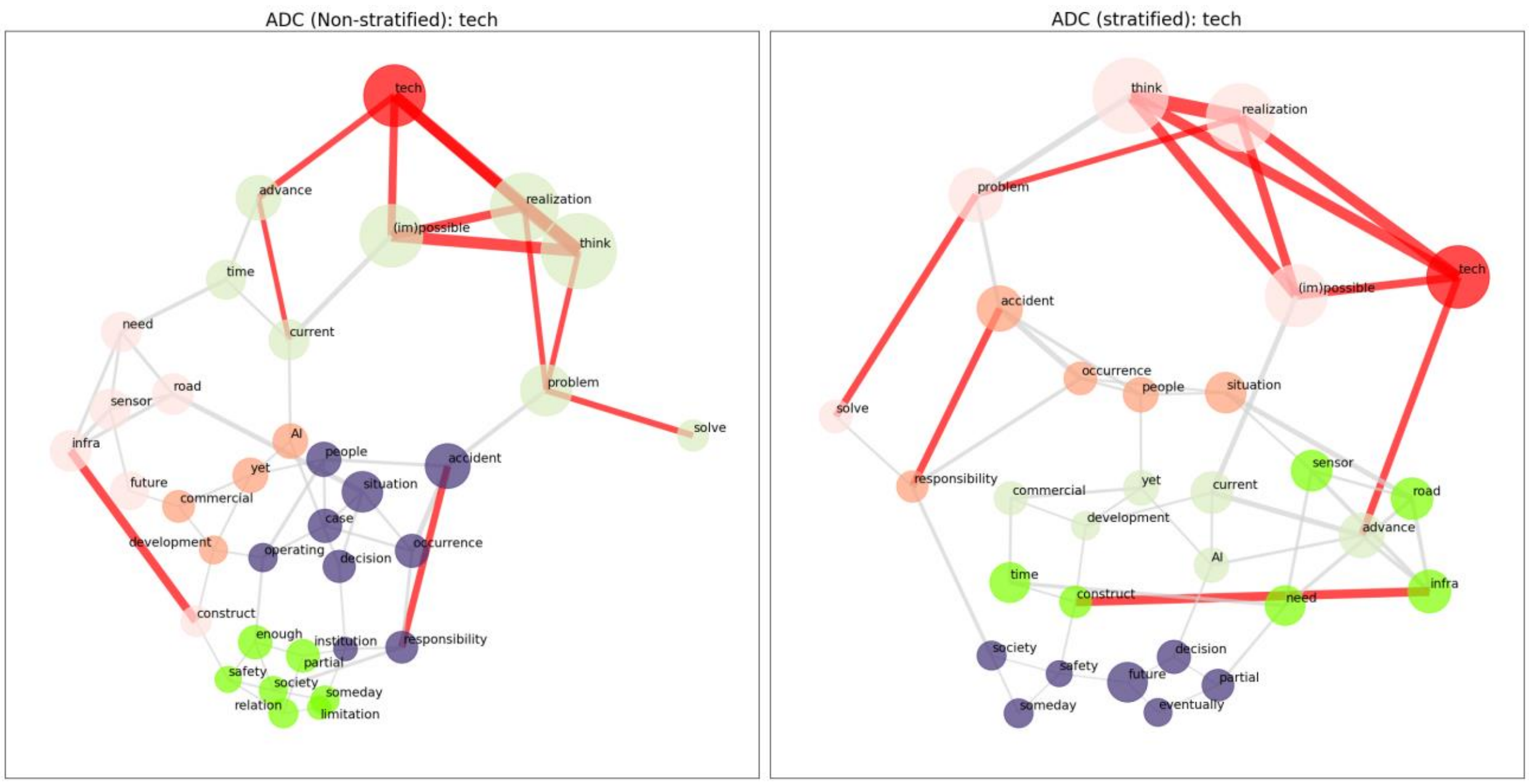

**Fig. 5.** The Co-occurrence Networks in ADC Cases

In Fig 4, “realization” has high correlations with “infra” (> 0.35), “sensor” of ADC (> 0.34), and “infra” has strong connections with “construct” in Fig 5, indicating the need to consider that infrastructure construction for autonomous driving is also a social challenge beyond mere tech. Additionally, some respondents seem to believe that AI, a base tech for object detection of autonomous driving, still has limitations, and time may be needed for transitioning to FSD. In Fig 5, “AI” has primary connections with “time,” and in Fig 4, “time” is in the top 10 DC terms with “realization” and “tech” (> 0.35) and has primary connections with “need” in Fig 5.

### *5.2.2. Co-occurrence Networks in ADR Cases*

In the CNA results for ADR, respondents generally seemed to consider the seriousness of accidents when ADR enters sidewalks. This is because in Fig 6, both “sidewalk” and “accident” are included in the upper ranks of each other's top 10 DC terms (> 0.46), and in Fig 7, “accident” is strongly connected with “occurrence.” This observation is interesting. As identified in Introduction, although a few expressed concern, Korea's policymakers generally did not take the possibility of accidents when changing ADR policy. However, this observation suggests that the users who pre-emptively understood the ADR policy based on their engineering literacy and the given texts (See Appendix A.) may instead consider the possibility of ADR accidents occurrence.

Moreover, respondents seem to have evaluated the attempt to set the speed limit of ADR to 15km/h. This is because, in Fig 6, “velocity” is included in the top 10 DC terms with “sidewalk” (> 0.38), and in Fig 7, it is primarily connected with “15”, “km,” “per hour,” and “limitation.” Especially as these terms in Fig 7 are primarily connected with “danger,” it seems that respondents evaluated the risks of the attempt.

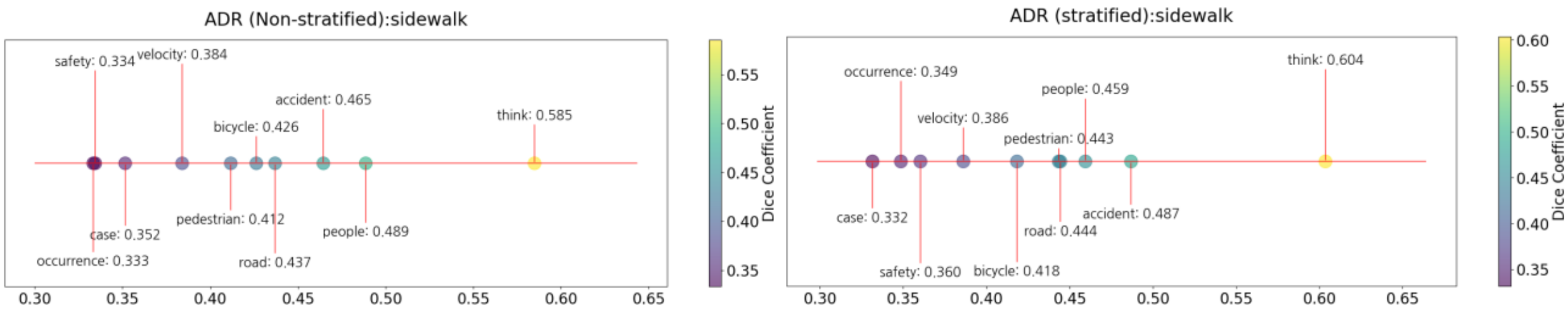

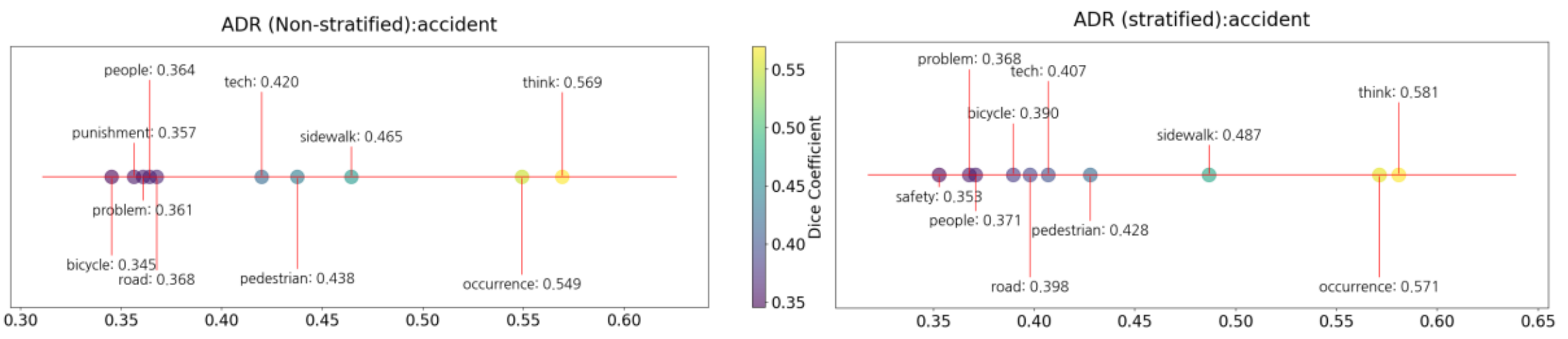

**Fig. 6.** The Top 10 neighboring Terms Ranked by Dice Coefficients in each ADR Case

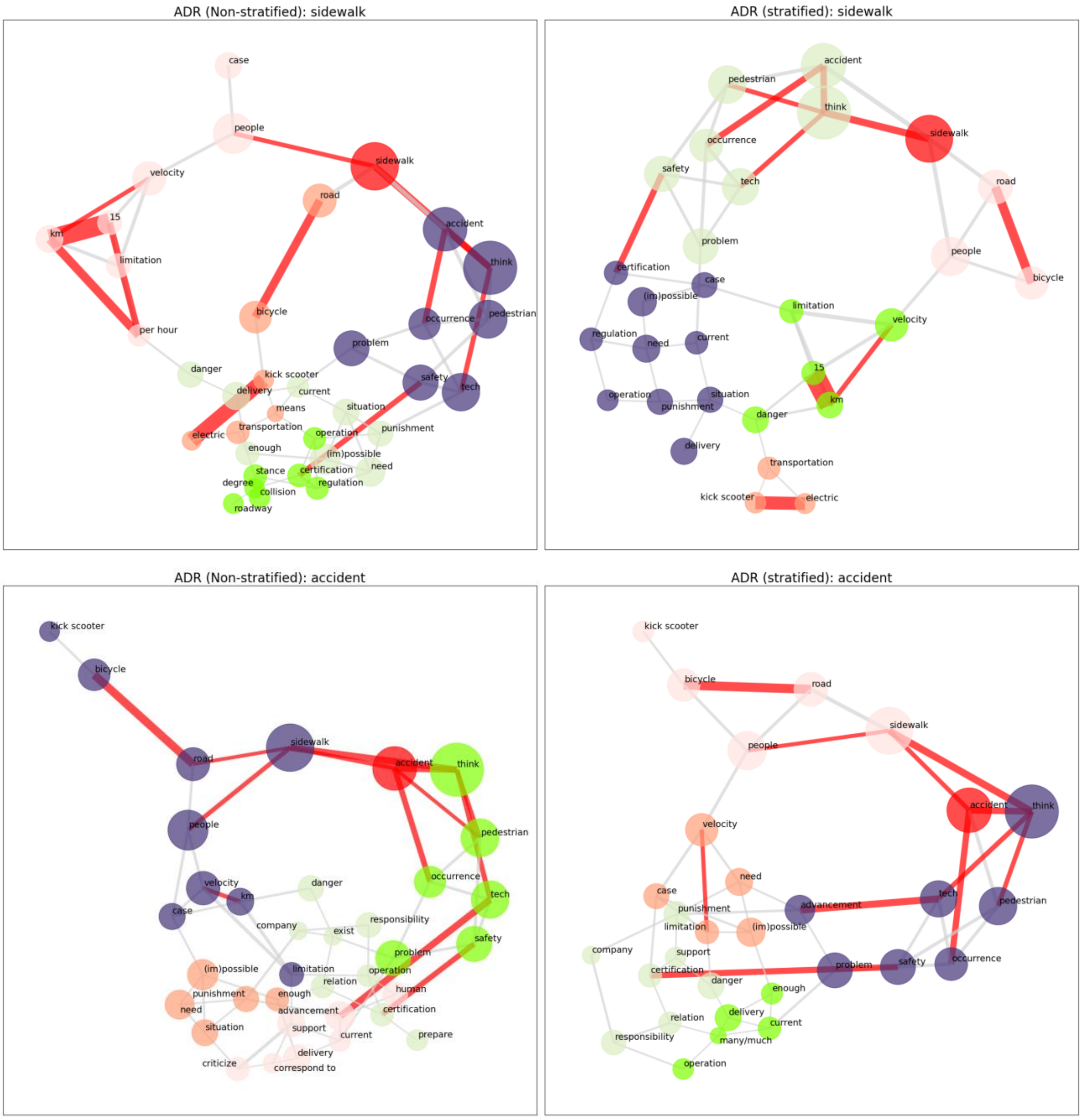

**Fig. 7.** The Co-occurrence Networks in ADR Cases

Some respondents seemed to consider the need for safety from ADR accidents on sidewalks or for operators or relevant companies to be appropriately punished in events of ADR accidents. In Fig 6, “safety” is included in the top 10 DC terms with “sidewalk” or “accident” ($> 0.33$), except when “accident” is the target keyword in non-stratified cases. In that excluded case, “punishment” replaces “safety” in the top 10 DC terms with “accident” ($\approx 0.357$), and in Fig 7, “punishment” is primarily connected with “company,” “operation,” and “need.” The ADR safety certification system and pedestrians' safety also seemed to be of primary concern to some respondents because in Fig 7, “safety” is strongly connected with “certification” and primarily with “pedestrian.”

Meanwhile, in Fig 6, “bicycle” is included in the mid-rank of the top 10 DC terms with “sidewalk” ($> 0.41$) and is strongly connected with “road” in Fig 7. In Korean, this collocation of “bicycle” and “road” refers to roads for the exclusive use of bicycles. Some respondents seemed to consider a compromise solution that separates ADR not only from cars but also from pedestrians.

### *5.3. Contextual Weights*

The CNA results were analyzed in the cautious form of guesses such as “It seems” because the collocations frequently appearing in the data cannot be guaranteed to be essential expressions contextually. However, Table 6 presents the Contextual Weights for keywords that represent the context of users' comments.

First, for ADC, respondents were generally optimistic about the feasibility of FSD. This is because “possible” ($> 92$) and “realization” ($> 62$) appear as keywords with the highest Contextual Weights across all the responses. Of course, the fact that “problem” ($> 51$) and “limitation” ($> 14$) are ranked third and eighth, respectively, reveals the recognition that there are still problems and limitations to transitioning to FSD. Specifically, “accident” ($> 43$), ranked fourth, was considered as a precondition for transitioning to FSD. “tech,” which was a target keyword in the CNA, is not included in Table 6. This is because, in the actual context, rather than “tech” itself, some aspects within that category were considered potentially influencing the transition of ADC to FSD. Keywords such as “V2V” ($> 10$), “traffic signal” ($> 12$), and “AI” ($> 11$) are examples of these elements.

Next, for ADR, “possible” ranks sixth in Table 6 ($> 29$) shows that some respondents believe solving these issues is possible. However, “safety,” which ranks fifth ($> 32$), is still emphasized as an essential value, and above all, “accident” ($> 83$) and “pedestrian” ($> 63$) rank at the top. The respondents generally considered potential ADR accidents and pedestrians. Despite the potential for ADR accidents being lightly dismissed as the ADR policy changed (See 1.), the actual users, who have knowledge about engineering techs and understood enough about the new policy (See Appendix A.), clearly treated this possibility as a contentious issue.

For exploring the users' optimism about the transition of ADC to FSD and concerns about the new ADR policy, “possible” and “realization” for ADC and “accident” and “pedestrian” for ADR were selected as target keywords.

**Table 6.** The Top 20 Keywords Ranked by Contextual Weights in Each Analysis Case

| Rank | ADC (Non-stratified) | | ADC (Stratified) | | ADR (Non-stratified) | | ADR (Stratified) | |
|---|---|---|---|---|---|---|---|---|
| | Word | Contextual Weight | Word | Contextual Weight | Word | Contextual Weight | Word | Contextual Weight |
| 1 | **<u>possible</u>** | **<u>128.8</u>** | **<u>possible</u>** | **<u>92.5</u>** | **<u>accident</u>** | **<u>106.2</u>** | **<u>accident</u>** | **<u>83.2</u>** |
| 2 | **<u>realization</u>** | **<u>88.7</u>** | **<u>realization</u>** | **<u>62.5</u>** | **<u>pedestrian</u>** | **<u>89.0</u>** | **<u>pedestrian</u>** | **<u>67.9</u>** |
| 3 | problem | 63.6 | problem | 51.3 | people | 71.6 | people | 49.5 |
| 4 | accident | 59.5 | accident | 43.6 | problem | 49.3 | problem | 39.0 |
| 5 | people | 31.1 | people | 23.2 | safety | 42.5 | safety | 32.2 |
| 6 | eventually | 22.9 | eventually | 18.5 | possible | 40.4 | possible | 29.3 |
| 7 | most | 19.1 | most | 14.3 | criticism | 36.9 | sidewalk | 26.7 |
| 8 | limitation | 16.2 | limitation | 14.0 | sidewalk | 34.5 | criticism | 23.6 |
| 9 | reason | 16.2 | traffic signal | 12.7 | km | 30.5 | km | 22.3 |
| 10 | traffic signal | 15.0 | prospect | 11.5 | per hour | 28.7 | kick scooter | 21.7 |
| 11 | AI | 15.0 | AI | 11.2 | kick scooter | 26.3 | per hour | 19.6 |
| 12 | artificial intelligence | 14.9 | V2V | 10.5 | electric | 18.1 | electric | 15.1 |
| 13 | situation | 14.8 | reason | 10.3 | roadway | 16.9 | punishment | 11.6 |
| 14 | prospect | 14.6 | situation | 9.9 | punishment | 16.8 | most | 11.2 |
| 15 | level | 13.8 | in the near future | 9.6 | eventually | 16.4 | harm | 11.2 |
| 16 | V2V | 12.8 | artificial intelligence | 9.3 | most | 15.2 | eventually | 11.1 |
| 17 | in the near future | 12.6 | someday | 9.1 | harm | 14.1 | roadway | 11.0 |
| 18 | prediction | 11.8 | level | 8.9 | electric wheelchair | 13.3 | electric wheelchair | 10.8 |
| 19 | environment | 11.5 | prediction | 8.9 | issue | 12.6 | argument | 8.2 |
| 20 | future | 11.4 | 100 | 8.1 | real | 10.6 | environment | 7.2 |

## *5.4. Contextual Semantic Networks*

Next, the terms with the top 10 cosine similarity values with each target keyword are presented in horizontal line graphs (See Fig 8, 10.), and more terms and edges are shown in contextual semantic networks. (See Fig 9, 11.)

Each network includes terms with cosine similarity values (CSV) within the top 4% with each target keyword, and their sizes reflect the contextual weights. Each node has edges up to the top 5 CSV terms that range from 0 to 1, and their thicknesses reflect the CSV. Some edges that reflect strong correlations within the top 20% of CSV in the corpus, including outside each network, were highlighted in red.

Incidentally, each target keyword was highlighted in red, and the colors of neighboring nodes were determined through the widely used Leiden Algorithm, which detects communities among nodes with high connectivity. (Traag et al., 2019) The layout of the networks was determined using the Fruchterman-Reingold Algorithm, which assumes the existence of springs acting as attraction or repulsion between nodes and adjusts the positions of nodes to minimize system energy. (Fruchterman and Reingold., 1991)

### *5.4.1. Contextual Semantic Networks: ADC*

In the C-SNA results for ADC, as seen in Fig 9, “realization” and “possible” are strongly connected with “problem” and “accident” to form a distinct community, despite having different meanings. This shows that while the respondents generally had optimistic views about transitioning to FSD, they recognized that problems with ADC accidents exist.

Meanwhile, in the stratified case where “realization” is the target keyword in Fig 9, unlike CNA, “responsibility” was omitted as it did not surpass the threshold. This does not imply that the issue of responsibility in ADC accidents was not significantly considered, but rather that “accident” and “responsibility” competed to represent the responses contextually during KeyBERT's filtering. While “responsibility” was not enough to overwhelm “accident” in the competition, it survives in the non-stratified case in Fig 9. However, in this case, the opinions of male respondents or those from the “industrial” or “mechanical” fields may have been more reflected. (See 4.1.)

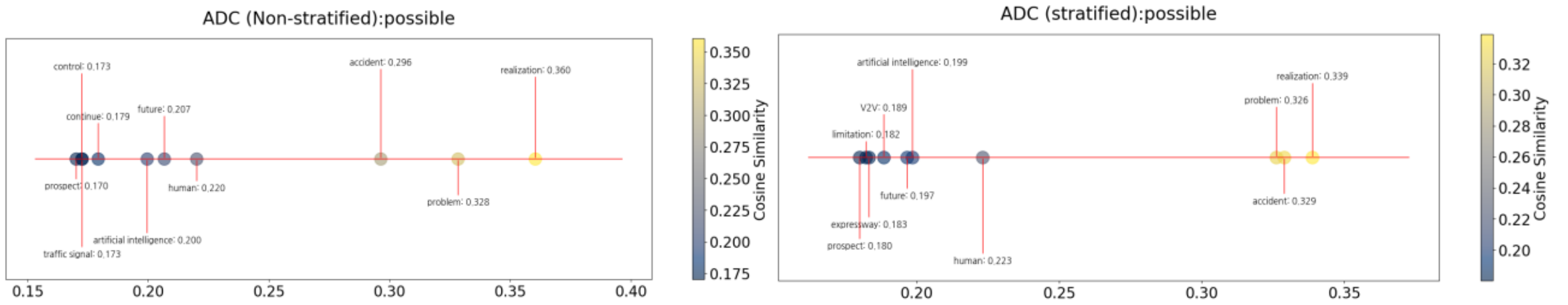

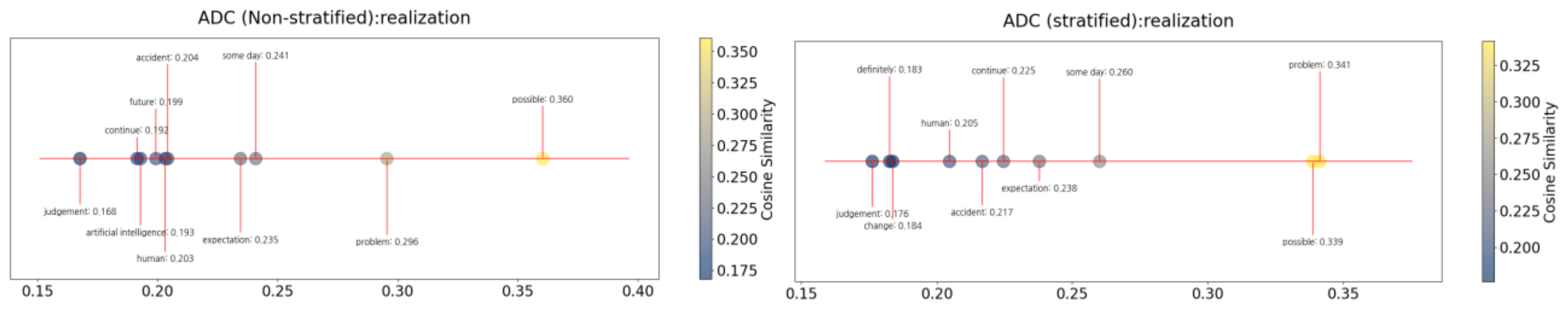

**Fig. 8.** The Top 10 Terms Ranked by Cosine Similarity Values in ADC Cases

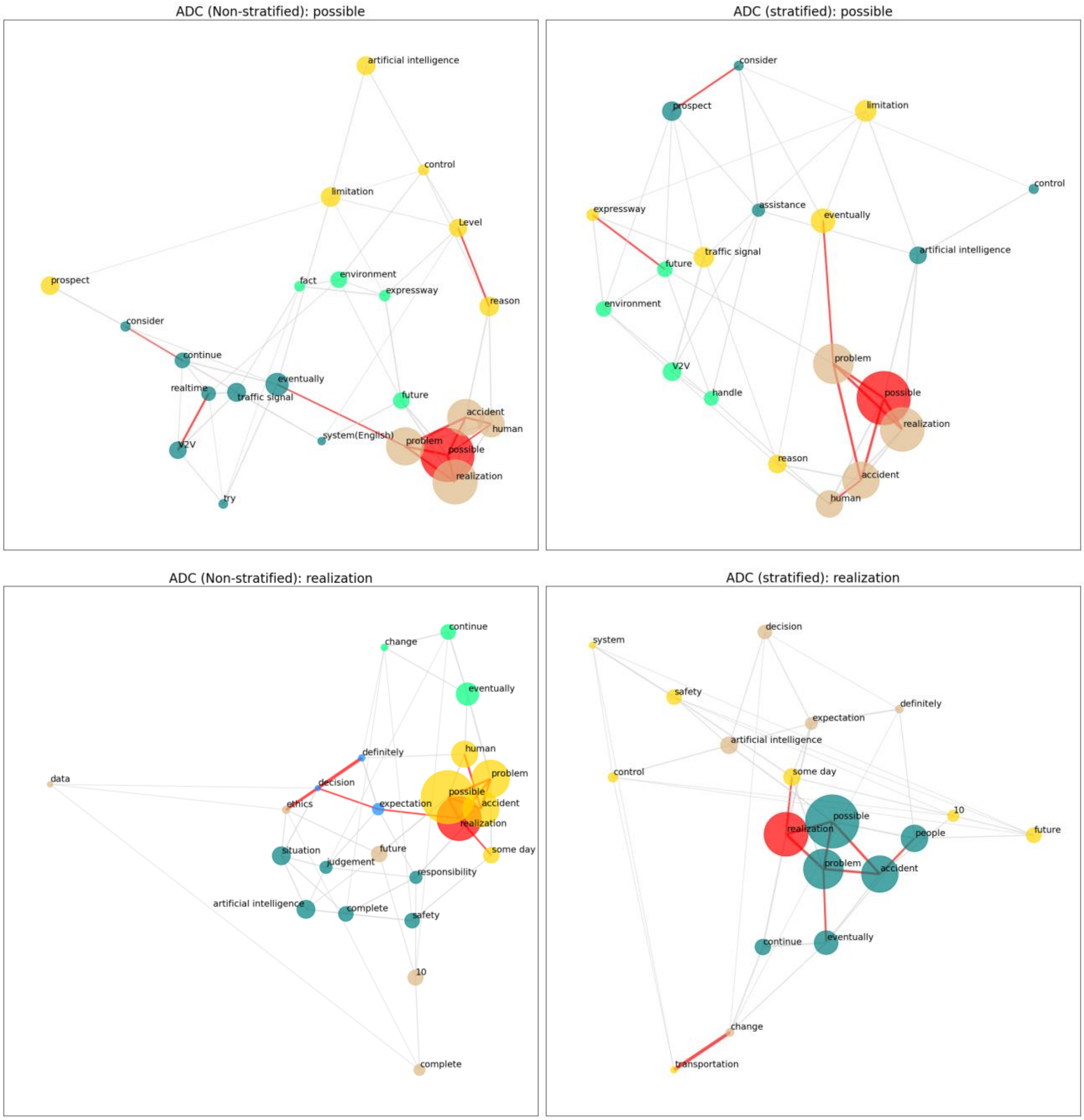

**Fig. 9.** The Contextual Semantic Networks in ADC Cases

Also, while "infra" and "construct" were strongly connected in CNA, this connection does not appear in Fig 9. This is because the effect of the infrastructure was more important than the infrastructure itself in the actual context. For example, some respondents expected FSD to become possible if V2V real-time communication becomes possible and limitations of sensors that cannot perfectly perceive traffic signals are compensated. In Fig 8, "V2V" ($\approx 0.189$) and "traffic signal" ($\approx 0.173$) are included in the top 10 CSV terms with "possible," and in Fig 9, "V2V" and "traffic signal" are primarily connected. However, some respondents recognized that there are still limitations in AI, a base tech for object detection. In the cases where "possible" is the target keyword in Fig 9, "artificial intelligence" is primarily connected with "limitation."

Finally, individual differences exist in the respondents' outlook on the timing for transitioning to FSD. In Fig 8, "realization" had a high CSV with "someday" ($> 0.24$), but it is not possible to specify the exact timing of "someday."

### *5.4.2. Contextual Semantic Networks: ADR*

In the C-SNA results for ADR, the respondents generally associated pedestrians with ADR accidents. This is because in Fig 10, "accident" and "pedestrian" are included in the highest ranks of each other's top 10 CSV terms. ($> 0.32$) Some respondents particularly emphasized pedestrians' safety as an essential value. "safety" is in the top 10 CSV terms with both "accident" and "pedestrian." ($> 0.17$) However, some respondents believed that the entry of ADR into sidewalks is, after all, something possible. "possible" is included in the top 10 CSV terms with both "accident" and "pedestrian" in Fig 10. ($> 0.17$)

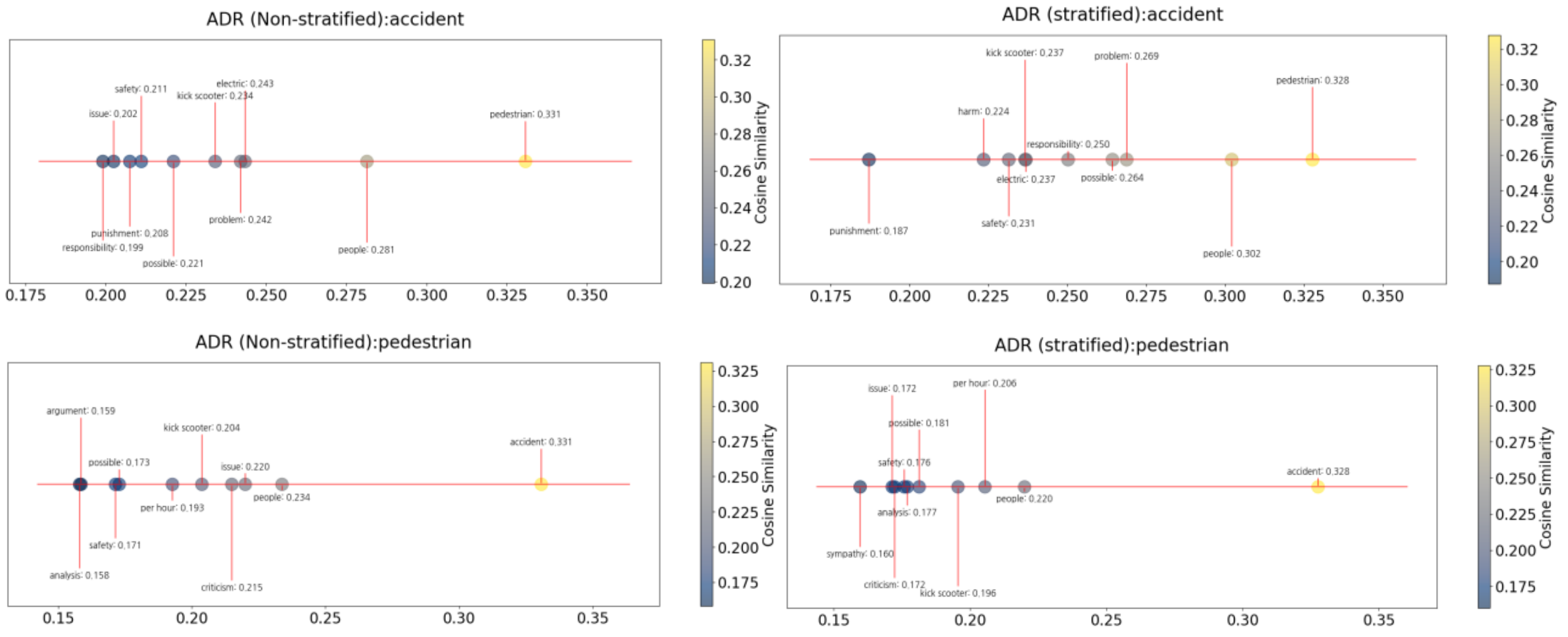

**Fig. 10.** The Top 10 Terms Ranked by Cosine Similarity Values in ADR Cases

The interesting point is that rather than taking clear stances between lenient and strict attitudes toward ADR, some respondents thought it possible to overcome such issues while also placing importance on ensuring pedestrians' safety from ADR accidents. In Fig 11, "possible" and "safety" nodes of similar size are strongly connected.

Opinions on how to ensure pedestrians' safety were various. On the one hand, the need to clarify responsibility in events of ADR accidents and to punish ADR operators appropriately was considered. In Fig 10, "responsibility" ($> 0.19$) and "punishment" ($> 0.18$) are included in the Top 10 CSV terms with "accident." On the other hand, interests in the policy to ensure the safety of ADR through safety certification systems were condensed into "safety."

Of course, while "safety" and "certification" were strongly connected in CNA (See Fig 7.), "certification" does not appear in Fig 11 as it did not surpass the threshold. This is because the Korean terms "안전 safety" and "인증 certification," which co-occurred in the responses, competed in KeyBERT's filtering to represent the responses contextually. In the non-stratified case where "accident" is the target keyword in Fig 11, the respondents used the compound word "안전인증 safety certification" more, and here "safety certification" survived.

In CNA, there was a theme about the need to construct roads for the exclusive use of ADR, like roads for the exclusive use of bicycles, to separate ADR and pedestrians, but this cannot be seen in Fig 11. This theme appears to have been less significant in the actual context of the responses. On the other hand, although not explicitly mentioned, "kick scooter" appeared in CNA (See Fig 7.) and is included in the top 10 CSV terms with "accident" and "pedestrian" in Fig 10, regardless of stratification. (> 0.19) And in Fig 11, "kick scooter" is strongly connected with "accident."

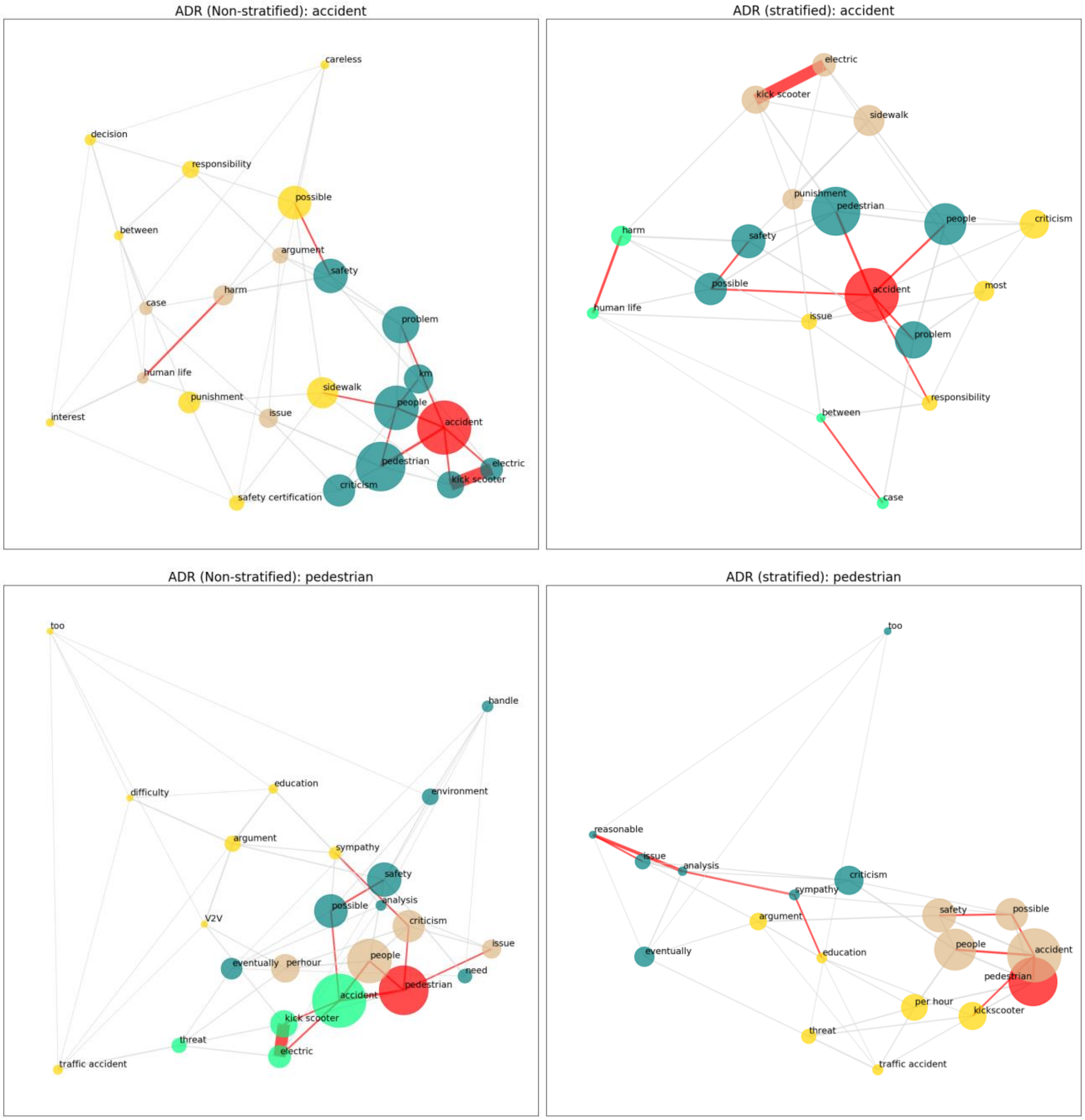

**Fig. 11.** The Contextual Semantic Networks in ADR Cases

There is a specific Korean context to note about kick scooters. In Korea, bicycles and kick scooters frequently encroach on sidewalks due to a lack of roads for the exclusive use of them. These actions are originally illegal. Especially with the proliferation of shared kick scooters, recurring accidents and kick scooters abandoned on sidewalks have significantly increased pedestrians' fatigue. As a result, the number of news articles about the need for regulations of shared kick scooter businesses is on the rise. (Kim, K. O. and Shin, Y. R., 2023) The appearance of “kick scooter” as a keyword representing users' comments about the ADR policy and its high CSV with “accident” and “pedestrian” is by no means a good signal for Korea's policymakers and ADR manufacturers.

This implies the potential risk that user fatigue with kick scooters could be projected onto ADR newly entering the sidewalks. Of course, the government takes a strategy claiming that ADR would be safe due to its slow speed. (National Assembly Secretariat, 2023b) However, some respondents with engineering literacy in this study thought that the limited speed of ADR could threaten pedestrians. In Fig 10, “per hour” is included in the top 10 CSV terms with “pedestrian” (> 0.19), regardless of stratification, and in Fig 11, “per hour” is primarily connected with “threat.”

## 6. Comparison and Summary

In this section, the efficiency of the CNA and C-SNA approaches conducted in Analysis and Results was compared, and the themes found were summarized. As consistent target keywords were not selected in both approaches, the efficiency was compared based on cases where the same target keywords were chosen. As shown in Table 7, Although more generous thresholds were set in C-SNA than CNA, fewer nodes and features were used in networks.

The reason is that KeyBERT was performed in advance, which extracts keywords to be used in C-SNA. KeyBERT only allowed keywords that could most contextually represent the responses to remain, reducing the number of features used in C-SNA to less than a third of those in CNA. (See Table 7.) Thanks to this condensation, interest in nodes could be saved and shifted to the correlations among terms, i.e., edges. This suggests that C-SNA can more efficiently represent the context of the text. The maximum number of edges per node was limited to 4 in CNA to avoid the visual complexity of the networks, but in C-SNA, it could be increased to 5.

**Table 7.** The Differences in thresholds: the number of nodes used, and features based on each SNA approach

| Topics | | ADC | | | | ADR | | | |
|---|---|---|---|---|---|---|---|---|---|
| Target keywords | | realization | | | | Accident | | | |
| Approaches | | CNA | | C-SNA | | CNA | | C-SNA | |
| Whether stratified or not (N/Y) | | N | Y | N | Y | N | Y | N | Y |
| Terms within the top α% as neighboring nodes | Cosine similarity | | | 4% | 4% | | | 4% | 4% |
| | Dice coefficient | 2% | 2% | | | 2% | 2% | | |
| Maximum number of edges per node | | 4 | 4 | 5 | 5 | 4 | 4 | 5 | 5 |
| The number of nodes used | | 35 | 30 | 23 | 19 | 36 | 30 | 23 | 18 |
| The number of features used | | 1703 | 1485 | 565 | 462 | 1753 | 1464 | 563 | 440 |

Such condensation of information, as in the case of “accident” and “responsibility,” to remain only the more comprehensive keyword “accident” (See 5.4.1.), induces little abstraction but does not bring significant loss to the information discernable from the networks. Aiming that each node of C-SNA represents the context of the text, appropriate interpretations can be compensated. Furthermore, in studies using text mining, using more than one method complementarily is common. As in this study, CNA and C-SNA can be performed together to gather the necessary information to understand the users' voices. The themes identified in Analysis and Results are summarized in Table 8. The information obtained from different approaches is similar; in some cases, C-SNA is more specific.

**RQ1.** What concerns about ADC transitioning to FSD can be observed from the users' responses?
**RQ2.** How do the users respond to the ADR policy change?

Based on the contents summarized in Table 8, RQ1 and RQ2 can be answered. In response to RQ1, the users were generally optimistic about the feasibility of ADC's FSD, but negative acceptance also appeared. They recognized solving ADC accident problems, including clarifying responsibility, as a precondition for transitioning to FSD. The opinions surfaced that FSD would be possible if improvements were made in sharing V2V traffic information and ADC's traffic signal perception capability through infra construction, as well as opinions that there are still limitations in AI, a base tech of ADC's object detection. Some users seemed to empathize more with Hyundai's vision of complementing ADC with V2V and V2I communications than Tesla's goal of pure vision-based FSD.

In response to RQ2, the users generally considered the likelihood of ADR accidents and the assurance of pedestrians' safety essential. However, some users think it is possible to overcome these issues. The requirements to ensure pedestrians' safety included clarifying responsibilities in case of accidents, the appropriate punishment for ADR operators, the realization of the safety certification system, and a speed limit that does not threaten pedestrians. Notably, the risk of negative user experiences with kick scooters being projected onto ADR was also explored.

**Table 8.** The Comparison of the Themes Using CNA and C-SNA Approaches

| Topic | ADC | |
|---|---|---|
| SNA Approaches | CNA | C-SNA |
| Themes | Contents | |
| Feasibility of FSD | It cannot be identified. (See Appendix B.) | Generally, the respondents think it is feasible but are aware of accident problems. |
| Accident | Solving accident problems (primarily responsibility) is significant. | Solving accident problems (including responsibility) is significant. |
| Tech | Advances in techs are essential to realizing FSD, but challenges are ahead. | The effects are more significant than the techs themselves. |
| Infra | Constructing infra to support ADC is essential. | It is expected to enable V2V, V2I communication and compensate for the limitations of traffic signal perception. |
| AI | Some respondents believe AI still has limitations. | Some respondents believe AI still has limitations. |
| FSD realization timing | It may need time. | There are individual differences in outlook. |
| **Topic** | **ADR** | |
| SNA Approaches | CNA | C-SNA |
| Themes | Contents | |
| Entering sidewalks | It cannot be identified. (See Appendix B.) | Generally, the respondents considered ADR accident problems and the need to ensure pedestrians' safety, but some believe it is possible to overcome these issues. |
| Accident and Safety | Generally, the respondents believe that accidents could occur and that pedestrians' safety should be ensured. Some are interested in the safety certification system. | Generally, the respondents believe that accidents could occur and that pedestrians' safety should be ensured. (Including interest in the safety certification system) |
| Responsibility and Punishment | Some considered the need for appropriate responsibility and punishment for relevant companies or ADR operators in the event of accidents. | Some considered the need for appropriate responsibility and punishment for relevant companies or ADR operators in the event of accidents. |
| Speed limit | Some evaluated the danger of the ADR speed limit. | Some considered the ADR speed limit and believe it may threaten pedestrians. |
| Roads for the exclusive use | Some believe there is a need to construct roads for the exclusive use of ADR, like bicycles, to separate ADR from pedestrians and cars. | It was a less significant theme. |
| The analogy with other moving objects | The potential risk of negative user experiences with kick scooters being projected onto ADR was explored. | The potential risk of negative user experiences with kick scooters being projected onto ADR was explored. |

## 7. Discussions

Some ADC manufacturers adopt a lowest liability risk design strategy. Under this strategy, by proactively defining rules of response for dangerous situations on the road, they can pass liability of accidents to drivers who fail to adhere to even one of these rules. (Martinho et al., 2022) Furthermore, the ADR policy change in South Korea has weakened the liability of ADR operators and manufacturers.

This reality subtly fractures with the concept of autonomous driving in the sense that not just corporations but users, too, should be free from accidents. The participants of this study generally considered solving AV accident problems a significant precondition for autonomous driving. Once AVs make a large-scale entry into public roads, it is only a matter of time before other user groups, less familiar with engineering tech than these engineering students, become aware that AV-related companies are avoiding accident liability.

Several fatal accidents can trigger users to be disappointed in ADC, and marketing that confuses the actual level of products, as in the case of Tesla, can further fuel this disappointment. The potential risk of negative user experiences with kick scooters being projected onto ADR has already been confirmed. As the role of policymakers is to assign appropriate levels of liabilities to AV-related companies (Anderson et al., 2016), policymakers need to provide the basis for users to feel systemically protected from AV accidents. If they fail to do this, like the sudden unintended acceleration issue already faced, similar issues could be reproduced in the case of AV.

According to Article 3-2 of the Product Liability Act in Korea, the injured person of the product defect must prove all the following three facts to presume that the product had a defect at the time the product was supplied and damage was caused because of the defect. (Korea Legislation Research Institute, 2019)

**A.** The Damage was caused to the injured person while the product was being used normally.
**B.** The damage referred to in A was attributable to a cause practically controllable by the manufacturer.
**C.** The damage referred to in A would not ordinarily be caused if it were not for the relevant defect of the product.

These conditions are more complicated than in the U.S., where case law has been formed to presume the manufacturer's strict liability once condition **A** is proven. (Ha, C. R and Kim, E. B., 2018) Korean courts have yet to presume the manufacturer's liability in a single case related to sudden unintended acceleration accidents.

Under such a policy, similar problems may reoccur in the case of AV. For instance, consider if the roles of ADC manufacturers and operators were separated in a car-sharing business. If an accident occurs because an ADC fails to perceive a traffic signal, it could be difficult for the victim to prove the liability of the relevant companies. The manufacturer could claim that the operator failed to perform regular inspections on the sensors, and the operator could claim that manufacturing defects were the cause of the accident, each shifting responsibility onto the other. With limited access to the companies' internal information for the victim, it could be challenging to prove **B**.

Even if a pedestrian dies after being hit by an ADR that rolled down from stairs, it could be challenging to prove **C** if the manufacturer claims that it was an exceptional case that could not be predicted at the time of supply. However, the manufacturer may also be at fault for providing an operator with an API that could trigger an ADR that did not perceive stairs. Even if the operator who carelessly operated the ADR near the stairs is at fault, the new ADR policy's milder penalties may not deter the operator.

The Korean government also claims that the speed limit of 15km/h for ADR, like the speed of electric wheelchairs, is safe (National Assembly Secretariat, 2023b) and that manufacturers can be appropriately controlled through a safety certification system. (National Assembly Secretariat, 2023c) However, allowing electric wheelchairs on sidewalks is based on social considerations for mobility-impaired persons, while ADR is based on corporate or individual benefits, which are different contexts. Some respondents with engineering literacy in this study also recognized the speed limit of 15km/h as potentially threatening to pedestrians.

The safety of the product itself and the safety within the socio-technical system are not always consistent. (Leveson, 2012) The certification system can define product specifications like speed limit, weight, and material. However, the desire of companies to handle more orders by bearing minor fines in various road environments can be challenging to control solely through product certification. Even if the criminal liability of AV-related companies is replaced with monetary compensation like AV insurance, the possibility that insurance burdens may be passed on to users in service fees or that insurance companies may choose civil litigation to avoid compensation payment should be considered.

Considering these current policy conditions, mass deployment of AV on public roads appears to be disadvantageous to users. This imbalance can slow users' acceptance of AV as innovative tech. Korean policymakers should change the policy to establish manufacturer liability once **A** is proven, as in the U.S., and strengthen the penalties to a more realistic level for accidents caused by ADR operators' culpabilities to protect users. Of course, technological advancement may be discouraged if only significant liability is imposed on AV-related companies. Therefore, it is necessary to construct V2V, V2I infra, or roads for the exclusive use of ADR to support AV safety enhancement, as suggested by some participants in this study.

## 8. Conclusions

In this study, contrary to Korean policymakers' underestimation, the fact that the participants generally regarded solving AV accident problems as a significant precondition for accepting AV was observed. While the users were generally optimistic about the feasibility of FSD for ADC, they showed interest in clarifying responsibility for ADC accidents and constructing V2V, V2I infra to support safe autonomous driving. Regarding the new policy that permits ADR on sidewalks by weakening related companies' liability, users generally are concerned about the potential for ADR accidents and pedestrians' safety. Nevertheless, some believe that overcoming these issues is possible.

This acceptance appeared among engineering students familiar with new tech, and other user groups less familiar with AV may take more stringent stances on future AV accident issues. To develop balanced AV policies, it is necessary to relax the burden of proving AV defects to a level like that of the U.S. and to strengthen the criminal liability of ADR operators to a realistic level.

In this study, users' voices were explored through CNA, a classic frequency-based SNA method, and C-SNA using KeyBERT, which only remains the keywords that represent the users' responses contextually. It demonstrated that in C-SNA, we could obtain information about user acceptance using fewer nodes and features than CNA. C-SNA is expected to complement diverse AV policy studies using text mining in the future.

Of course, this study has limitations. While quantitative indicators like TF-IWF, Dice coefficient, and cosine similarity were used to control the researchers' heuristics in extracting semantic networks, the researchers were left to interpret the extracted results. However, it is common in text mining for the interpretation of results to depend on the researchers' domain knowledge, and text mining is primarily used for exploratory research rather than for proving causality. The collected responses were fewer than 400, which allowed the researchers to read all the responses and interpret the SNA results, trying to minimize the possibility of misunderstanding users' voices.

For future works, based on the AV user acceptance explored in this study, it is expected that a more rigorous AV policy study can be conducted by modifying the UTAUT (Unified Theory of Acceptance and Use of Technology) model (Venkatesh et al., 2003) and statistical testing the model.

## Funding

This work is supported by Basic Science Research Program through the National Research Foundation of Korea (NRF) funded by the Ministry of Education (NRF-2017R1D1A1B05029080).

## Acknowledgement

This work was approved by the IRB of Soongsil University (SSU-202202-HR-457-1). We are deeply grateful to Yongmin Yoo, a Ph.D. candidate in Computer Science at the University of Auckland, who recommended using KeyBERT as the primary method for this study. Any shortcomings in this study are the sole responsibility of us who applied the methods, not him. This study would not have been possible without the participation of graduate students from the following universities: Konkuk University, Kyung Hee University, Korea University, KwangWoon University, Kookmin University, Duksung Women's University, Dongguk University, Dongduk Women's University, Sangmyung University, Sogang University, Seoul National University of Science and Technology, Seoul National University, University of Seoul, Seoul Women's University, Sungkyunkwan University, Sungshin Women’s University, Sejong University, Sookmyung Women's University, Soongsil University, Yonsei University, Ehwa

Womans University, Chung-Ang University, Hankuk University of Foreign Studies, Hansung University, Hanyang University, Hongik University.

## CRediT authorship contribution statement

**Jinwoo Ha:** Conceptualization, Methodology, Software, Formal analysis, Investigation, Writing - Original Draft, Writing - Review & Editing. **Dongsoo Kim:** Validation, Writing - Review & Editing, Supervision, Project administration, Funding acquisition.

## Declaration of Generative AI and AI-assisted technologies in the writing process

During the preparation of this work, the first author used ChatGPT, DeepL, and Grammarly to improve readability and language. After using these services, the author reviewed and edited the content as needed and takes full responsibility for the content of the publication.

## Data availability

Conditionally available. The data for this study was collected through the survey. Therefore, by human subject research ethics, any requester and data provider must receive approval for data reuse from their respective Institutional Review Boards (IRB).

## Appendix A. Questionnaire

Refer to the following. Detailed items, such as questions to identify sociodemographic information, were included in the questionnaire but omitted in this paper due to space limitations.

**[Given text 1] Please read the following and answer the question.**

① Bad news for autonomous driving: Tesla, the self-driving car manufacturer, had promised that autonomous driving could self-drive with cameras and software upgrades. However, as the company added radar sensors to its new cars, some consumers accused it of exaggerated advertising. The California State Legislature in the U.S. passed a law prohibiting Tesla from advertising partially autonomous driving as full self-driving, judging Tesla was misleading consumers. The National Highway Traffic Safety Administration in the U.S. is investigating a November 2022 eight-vehicle pileup in San Francisco caused by a sudden stop by a Tesla vehicle and confirmed that the vehicle had activated autonomous driving shortly before the crash.

② Good news for autonomous driving: Hyundai-Kia Motors Group has been piloting Level 4 self-driving taxis on 26 roads (48.8 km) in Gangnam, Seoul, and has constructed infra with the Seoul Metropolitan Government to link traffic signals to autonomous vehicles, considering that it is difficult for vehicles to perceive 100% of traffic signals using only sensors and that it is necessary to prepare for sensor failures. The company plans to commercialize Level 3 self-driving cars on highways in South Korea and Level 4 robo-taxis in North America. Meanwhile, the Ministry of Land, Infrastructure and Transport plans to accelerate the commercialization of autonomous driving by constructing communication infra on roads nationwide that enable real-time V2V (Vehicle to Vehicle) and V2R (Vehicle to Road) information sharing.

Q4. Regarding [Given text 1], what is your outlook on whether FSD will be realized?

**[Given text 2] Please read the following and answer the question.**

Autonomous driving robots are the next generation of transport, promising to reduce labor costs. The robots were classified as “vehicles” and could not drive on sidewalks without a permit for a while. Starting this year, they will be included in the scope of pedestrians and can drive on sidewalks. The following issues exist with this action.

① A critical stance: Legally, a traffic accident means an accident caused by vehicle traffic. Including robots as pedestrians make it harder to punish robot operators on sidewalks for injuring or causing the death of people. The legislative intent is to protect pedestrians on sidewalks, and it does not seem reasonable to consider robots as pedestrians to activate them.

② An advocacy stance: It is not like no penalties are in place. Suppose a robot operator operates a robot in a way that causes danger and impediment to a pedestrian. In that case, the operator will be punished with a fine not exceeding 200,000 won, misdemeanor imprisonment, or a minor fine. The limit speed will also be defined as 15km/h, which is about the same as electric wheelchairs, through the enforcement decree, so it should not be too dangerous. If there is an accident, private insurance can compensate.

③ A re-critical stance: Punishment is a matter of the government's interpretation, and the government's default stance is that accidents caused by robots are treated as accidents between pedestrians. If a human walks at 5km/h and runs at 10km/h, 15km/h seems fast, and infants can be hard to avoid. Electric wheelchairs are operated by humans anyway, so it is a concern if robots that get around on their own can reach speeds of up to 15km/h.

④ A re-advocacy stance: Only safety-certified robots will be allowed on sidewalks. No accidents have occurred during the autonomous robots testing period, and the public has responded favorably. Since the robots are targeting the short-distance delivery market, they will never compete with delivery scooters on speed. As driving data accumulates, so autonomous driving algorithms will develop. It is hard to grow the tech when we burden it by heavily penalizing companies for accidents that might happen occasionally.

Q5. What is your opinion on the issue presented in [Given text 2]?

## Appendix B. Why was the “가능” translated differently depending on the analysis methods?

In Korean, the prefix “불 im” can combine with the free morpheme “가능 possible” to form “불가능 impossible.” Since this study only used free morphemes for analysis, it is impossible to infer in frequency-based text mining whether the original form of “가능” was “가능 possible” or “불가능 impossible.” Therefore, “가능” was translated with reserve as “(im)possible” in CNA. However, in C-SNA, “가능” was translated as “possible.”

Because KoBigBird, adopted in the KeyBERT process, performs its subwords-based tokenizing before document embedding. (Park, J. W. and Kim, D. G., 2021) Therefore, KoBigBird recognizes “가능” and “불가능” as separate tokens and performs document embedding. Even only “가능” remains by removing “불” in the preparation stage for token embedding; if the context of a document is closer to “불가능 impossible,” the document embedding is hard to have a high cosine similarity value to the embedding of “가능 possible.” In other words, if “가능” was extracted as a keyword, it suggests that the context of the document was closer to “possible.”

## References

[1] Anderson, J. M., Nidhi, K., Stanley, K. D., Sorensen, P., Samaras, C., & Oluwatola, O. A. (2016). *Autonomous Vehicle Technology: A Guide for Policymakers*. Rand Corporation. https://doi.org/10.7249/rr443-2

[2] Jones, R., Sadowski, J., Dowling, R., Worrall, S., Tomitsch, M., & Nebot, E. (2023). Beyond the Driverless Car: A Typology of Forms and Functions for Autonomous Mobility. *Applied Mobilities*, *8*(1), 26-46. https://doi.org/10.1080/23800127.2021.1992841

[3] Li, J., Rombaut, E., & Vanhaverbeke, L. (2021). A systematic review of agent-based models for autonomous vehicles in urban mobility and logistics: Possibilities for integrated simulation models. *Computers, Environment and Urban Systems*, *89*, 101686. https://doi.org/10.1016/j.compenvurbsys.2021.101686

[4] Das, S., Dutta, A., Lindheimer, T., Jalayer, M., & Elgart, Z. (2019). YouTube as a Source of Information in Understanding Autonomous Vehicle Consumers: Natural Language Processing study. *Transportation Research Record*, *2673*(8), 242-253. https://doi.org/10.1177/0361198119842110

[5] Kohl, C., Mostafa, D., Böhm, & M., Krcmar, H. (Feb 2017). Disruption of Individual Mobility Ahead? A Longitudinal Study of Risk and Benefit Perceptions of Self-driving Cars on Twitter. In *Proceedings of 13, Internationale Tagung Wirtschafts informatik (WI 2017)*, 1220-1234. https://aisel.aisnet.org/cgi/viewcontent.cgi?article=1022&context=wi2017

[6] Io, H. N., & Lee, C. B. (December 2019). What are the Sentiments About the Autonomous Delivery Robots?. In *2019 IEEE International Conference on Industrial Engineering and Engineering Management (IEEM)* (pp. 50-53). IEEE. https://doi.org/10.1109/ieem44572.2019.8978921

[7] Bagloee, S. A., Tavana, M., Asadi, M., & Oliver, T. (2016). Autonomous vehicles: challenges, opportunities, and future implications for transportation policies. *Journal of Modern Transportation*, *24*(4), 284-303. https://doi.org/10.1007/s40534-016-0117-3

[8] Uhlemann, E., (2015). Introducing Connected Vehicles [Connected Vehicles]. *IEEE Vehicular Technology Magazine*, *10*(1), 23-31. https://doi.org/10.1109/mvt.2015.2390920

[9] Premebida, C., Serra, P., Asvadi, A., Valejo, A., & Moura, L. (June 2018). Cooperative ITS challenges: AUTOCITS Pilot in Lisbon. In *2018 IEEE 87th Vehicular Technology Conference (VTC Spring)* (pp. 1-5). IEEE. https://doi.org/10.1109/vtcspring.2018.8417881

[10] Naranjo, J. E., Jiménez, F., Gómez, A., Valle, A., González, J., & Cruz, A. (October 2020). Integration of C-ITS in Autonomous Driving. In *2020 IEEE Intelligent Vehicles Symposium (IV)* (pp. 27-32). IEEE. https://doi.org/10.1109/iv47402.2020.9304727

[11] Xing, Y., Boyle, L. N., Sadun, R., Lee, J. D., Shaer, O., & Kun, A. (2023). Perceptions related to engaging in non-driving activities in an automated vehicle while commuting: A text mining approach. *Transportation Research Part F: Traffic Psychology and Behaviour*, *94*, 305-320. https://doi.org/10.1016/j.trf.2023.01.015

[12] Dos Santos, F. L. M., Duboz, A., Grosso, M., Raposo, M. A., Krause, J., Mourtzouchou, A., … & Ciuffo, B. (2022). An acceptance divergence? Media, citizens and policy perspectives on autonomous cars in the European Union. *Transportation Research Part A: Policy and Practice*, *158*, 224-238. https://doi.org/10.1016/j.tra.2022.02.013

[13] Ding, Y., Korolov, R., Wallace, W. A., & Wang, X. C. (2021). How are sentiments on autonomous vehicles influenced? An analysis using Twitter feeds. *Transportation Research Part C: Emerging Technologies*, *131*, 103356. https://doi.org/10.1016/j.trc.2021.103356

[14] Li, T., Lin, L., Choi, M., Fu, K., Gong, S., & Wang, J. (November 2018). YouTube Av 50k: An Annotated Corpus for Comments in Autonomous Vehicles. In *2018 International Joint Symposium on Artificial Intelligence and Natural Language Processing (iSAI-NLP)* (pp. 1-5). IEEE. https://doi.org/10.1109/isai-nlp.2018.8692799

[15] Spärck Jones, K. (1972). A statistical interpretation of term specificity and its application in retrieval. *Journal of Documentation*, *28*(1), 11-21. https://doi.org/10.1108/eb026526

[16] Blei, D. M., Ng, A. Y., & Jordan, M. I. (2003). Latent Dirichlet Allocation. *Journal of Machine Learning Research*, *3*(Jan), 993-1022. https://dl.acm.org/doi/10.5555/944919.944937

[17] Das, S. (2021). Autonomous vehicle safety: Understanding perceptions of pedestrians and bicyclists. *Transportation Research part F: Traffic Psychology and Behaviour*, *81*, 41-54. https://doi.org/10.1016/j.trf.2021.04.018

[18] Chaney, A., & Blei, M. D. (May 2012). Visualizing Topic Models. In *Proceedings of the International AAAI Conference on Weblogs and Social Media* (Vol 6, No. 1, pp. 419-422). https://doi.org/10.1609/icwsm.v6i1.14321

[19] Sridhar, V. K. R. (June 2015). Unsupervised Topic Modeling for Short Texts Using Distributed Representations of Words. In *Proceedings of the 1st Workshop on Vector Space Modeling for Natural Language Processing* (pp. 192-200). https://doi.org/10.3115/v1/w15-1526

[20] Suominen, A., & Toivanen, H. (2016). Map of Science with Topic Modeling: Comparison of Unsupervised Learning and Human-assigned Subject Classification. *Journal of the Association for Information Science and Technology*, *67*(10), 2464-2476. https://doi.org/10.1002/asi.23596

[21] Harris, Z. S. (1954). Distributional Structure. *Word*, *10*(2-3), 146-162. https://doi.org/10.1080/00437956.1954.11659520

[22] Vaswani, A., Shazeer, N., Parmar, N., Uszkoreit, J., Jones, L., Gomez, A. N., … & Polosukhin, I. (December 2017). Attention Is All You Need. In *Proceedings of the 31st International Conference on Neural Information Processing Systems (NIPS 2017)* (pp. 6000-6010). https://dl.acm.org/doi/10.5555/3295222.3295349

[23] Devlin, J., Chang, M. W., Lee, K., & Toutanova, K. (June 2019). Bert: Pre-training of Deep Bidirectional Transformers for Language Understanding. In *Proceedings of the 2019 Conference of the North American Chapter of the Association for Computational Linguistics: Human Language Technologies* (Vol 1, pp. 4171-4186). http://dx.doi.org/10.18653/v1/N19-1423

[24] Lee, J. B. (2020). Kcbert: Korean comments BERT. In *Proceedings of the 32nd Annual Conference on Human and Language Technology* (pp. 437-440). https://koreascience.kr/article/CFKO202030060691828.page

[25] Lee, S. A., Jang, H. S., Baik, Y. M., Park, S. Z., & Shin, H. P. (2020). Kr-bert: A Small-Scale Korean-Specific Language Model. *arXiv preprint arXiv:2008.03979*. https://doi.org/10.48550/arXiv.2008.03979

[26] Lee, H. J., Yoon, J. W., Hwang, B. G., Joe, S. H., Min, S. J., & Gwon, Y. J. (January 2021). Korealbert: Pretraining a Lite Bert Model for Korean Language Understanding. In *2020 25th International Conference on Pattern Recognition (ICPR)* (pp. 5551-5557). IEEE. https://doi.org/10.1109/icpr48806.2021.9412023

[27] Park, S. J., Moon, J. H., Kim, S. D., Cho, W. I., Han, J. Y., Park, J. W., … & Cho, K. (2021). Klue: Korean Language Understanding Evaluation. *arXiv preprint arXiv:2105.09680*. https://doi.org/10.48550/arXiv.2105.09680

[28] Bouma, G. (2009). Normalized (Pointwise) Mutual Information in Collocation Extraction. In *Proceedings of the Biennial GSCL Conference* (Vol. 30, pp. 31-40). https://svn.spraakdata.gu.se/repos/gerlof/pub/www/Docs/npmi-pfd.pdf

[29] Lau, J. H., Newman, D., & Baldwin, T. (April 2014). Machine Reading Tea Leaves: Automatically Evaluating Topic Coherence and Topic Model Quality. In *Proceedings of the 14th Conference of the European Chapter of the Association for Computational Linguistics* (pp. 530-539). http://dx.doi.org/10.3115/v1/E14-1056

[30] Dieng, A. B., Ruiz, F. J. R., & Blei, D. M. (2020). Topic Modeling in Embedding Spaces. *Transactions of the Association for Computational Linguistics*, *8*, 439-453. https://doi.org/10.1162/tacl_a_00325

[31] Grootendorst, M. (2022a). BERTopic: Neural topic modeling with a class-based TF-IDF procedure. *arXiv preprint arXiv:2203.05794*. https://doi.org/10.48550/arXiv.2203.05794

[32] McInnes, L., Healy, J., & Melville, J. (2018). Uniform Manifold Approximation and Projection for Dimension Reduction. *arXiv preprint arXiv:1802.03426*. https://doi.org/10.48550/arXiv.1802.03426

[33] Sindi, S., & Woodman, R. (2021). Implementing commercial autonomous road haulage in freight operations: An industry perspective. *Transportation Research Part A: Policy and Practice*, *152*, 235-253. https://doi.org/10.1016/j.tra.2021.08.003

[34] Martinho, A., Herber, N., Kroesen, M., & Chorus, C., 2021. Ethical issues in focus by the autonomous vehicles industry. *Transport Reviews*, *41*(5), 556-577. https://doi.org/10.1080/01441647.2020.1862355

[35] Hilgarter, K., & Granig, P. (2020). Public perception of autonomous vehicles: A qualitative study based on interviews after riding an autonomous shuttle. *Transportation Research Part F: Traffic psychology and behaviour*, *72*, 226-243. https://doi.org/10.1016/j.trf.2020.05.012

[36] Merfeld, K., Wilhelms, M. P., & Henkel, S., 2019. Being driven autonomously – A qualitative study to elicit consumers' overarching motivational structures. *Transportation Research Part C: Emerging Technologies*, *107*, 229-247. https://doi.org/10.1016/j.trc.2019.08.007

[37] Wang, X., Yang, L., Wang, D., & Zhen, L. (2013). Improved TF-IDF Keyword Extraction Algorithm. *Computer Science and Application*, *3*(1), 64-68. doi:10.12677/csa.2013.31012

[38] Tian, H., & Wu, L. (2018). Microblog Emotional Analysis Based on TF-IWF weighted Word2vec model. In: *2018 IEEE 9th International Conference on Software Engineering and Service Science (ICSESS)* (pp. 893-896). IEEE. https://doi.org/10.1109/ICSESS.2018.8663837

[39] Lu, X., & Jin, J., 2022. A study on the lists of common Korean stopwords for text mining. *Korean Language Research*, *63*(13), 1-15. http://doi.org/10.16876/klrc.2022.63.13.1

[40] Dice, L. R. (1945). Measures of the Amount of Ecologic Association Between Species. *Ecology*, *26*(3), 297-302. https://doi.org/10.2307/1932409

[41] Rychlý, P. (December 2008). A Lexicographer-Friendly Association Score. In *Proceedings of Recent Advances in Slavonic Natural Language Processing (RASLAN)* (pp. 6-9). https://www.fi.muni.cz/usr/sojka/download/raslan2008/13.pdf

[42] Kolesnikova, O. (2016). Survey of word co-occurrence measures for collocation detection. *Computación y Sistemas*, *20*(3), 327-344. https://doi.org/10.13053/cys-20-3-2456

[43] Pedregosa, F., Varoquaux, G., Gramfort, A., Michel, V., Thirion, B., Grisel, O., … & Duchesnay, É. (2011). Scikit-learn: Machine learning in Python. *The Journal of Machine Learning Research*, *12*, 2825-2830. https://dl.acm.org/doi/10.5555/1953048.2078195

[44] Zaheer, M., Guruganesh, G., Dubey, K. A., Ainslie, J., Alberti, C., Ontanon, S., … & Ahmed, A. (December 2020). Big bird: Transformers for Longer Sequences. In *Proceedings of 34th Conference on Neural Information Processing Systems (NeurIPS 2020)* (pp. 17283-17297). https://dl.acm.org/doi/abs/10.5555/3495724.3497174

[45] Bird, S., Ewan, K., & Edward, L. (2009). *Natural Language Processing with Python*. O'Reilly Media Inc. https://dl.acm.org/doi/10.5555/1717171

[46] Yoo, W. J. (2022). Introduction to Deep Learning for Natural Language Processing, Wikidocs. https://wikidocs.net/book/2155

[47] Park, E. L., & Cho, S. Z. (October 2014). KoNLPy: Korean natural language processing in Python. In: *Proceedings of the 26th Annual Conference on Human & Cognitive Language Technology* (Vol. 6, pp. 133-136). https://buildmedia.readthedocs.org/media/pdf/konlpy/stable/konlpy.pdf

[48] Park, K. B., Lee, J. H., Jang, S. B., & Jung, D. W. (2020). An Empirical Study of Tokenization Strategies for Various Korean NLP Tasks. arXiv preprint arXiv:2010.02534. https://doi.org/10.48550/arXiv.2010.02534

[49] Kudo, T., Yamamoto, K., & Matsumoto, Yuji. (2004). Applying Conditional Random Fields to Japanese Morphological Analysis. In: *Proceedings of the 2004 Conference on Empirical Methods in Natural Language Processing (EMNLP 2004)* (pp. 230-237). https://aclanthology.org/W04-3230/

[50] Vijayarani, S., Ilamathi, J., & Nithya. (2015). Preprocessing Techniques for Text Mining - An Overview. *International Journal of Computer Science & Communication Networks*, *5*(1), 7-16. https://www.researchgate.net/publication/339529230_Preprocessing_Techniques_for_Text_Mining_-_An_Overview

[51] Chang, T. Y., Chi, S. H., & Im, S. B. (2022). Understanding User Experience and Satisfaction with Urban Infrastructure through Text Mining of Civil Complaint Data. *Journal of Construction Engineering and Management*, *148*(8), 04022061. https://doi.org/10.1061/(ASCE)CO.1943-7862.0002308

[52] Park, S. T., & Liu, C. (2022). A study on topic models using LDA and Word2Vec in travel route recommendation: focus on convergence travel and tours reviews. Personal and Ubiquitous Computing, 26, 429-445. https://doi.org/10.1007/s00779-020-01476-2

[53] Lee, D. J., Yeon, J. H., Hwang, I. B., & Lee, S. G. (2010). KKMA: A Tool for Utilizing Sejong Corpus based on Relational Database. *KIISE Transactions on Computing Practices*, *16*(11), 1046-1050. https://koreascience.kr/article/JAKO201014435574668.page

[54] Traag, V. A., Waltman, L., & Van Eck, N. J. (2019). From Louvain to Leiden: guaranteeing well-connected communities. *Scientific Reports*, *9*(1), 5233. https://doi.org/10.1038/s41598-019-41695-z

[55] Kamada, T., & Kawai, S. (1989). An algorithm for drawing general undirected graphs. *Information Processing Letters*, *31*(1), 7-15. https://doi.org/10.1016/0020-0190(89)90102-6

[56] Fruchterman, T. M. J., & Reingold, E. M. (1991). Graph drawing by force-directed placement. *Software: Practice and Experience*, *21*(11), 1129-1164. https://doi.org/10.1002/spe.4380211102

[57] Kim, K. O., & Shin, Y. R. (2023). Analysis of Social Trends for Electric Scooters Using Dynamic Topic Modeling and Sentiment Analysis. *KIPS Transactions on Software and Data Engineering*, *12*(1), 19-30. https://doi.org/10.3745/KTSDE.2023.12.1.19

[58] Ha, C. R., & Kim, E. B. (2018). A Comparative Study on the Burden of proof between Korea and the USA under the Product Liability. *Korea Trade Review*, *43*(3), 101-124. https://koreascience.kr/article/JAKO201818053120363.page

[59] Leveson, N. G. (2012). *Engineering a Safer World: Systems Thinking Applied to Safety*. The MIT Press. https://doi.org/10.7551/mitpress/8179.001.0001

[60] Venkatesh, V., Morris, M. G., Davis, G. B., & Davis, F. D. (2003). User Acceptance of Information Technology: Toward a Unified View. *MIS Quarterly*, *27*(3), 425-478. https://doi.org/10.2307/30036540

## Web References

[1] Ministry of Land, Infrastructure and Transport. (13 September 2022). Never stop innovating for the future: the mobility innovating roadmap. Last accessed: 30 April 2023. http://www.molit.go.kr/USR/NEWS/m_71/dtl.jsp?lcmspage=1&id=95087208

[2] Park, K. I. (9 June 2022). Self-driving car "Roboride" drives around Gangnam. irobotnews. Last accessed: 30 April 2023. http://www.irobotnews.com/news/articleView.html?idxno=28697

[3] Kim, C. S. (1 March 2023). [New Year's Message] Chung Eui-sun, Hyundai Motor Group Chairman: "Need to build trust through endless challenges". MoneyS. Last accessed: 30 April 2023. https://moneys.mt.co.kr/news/mwView.php?no=2023010309414236293

[4] Kolodny, L. (25 May 2021). Tesla is ditching radar, will rely on cameras for Autopilot in some cars. CNBC. Last accessed: 30 April 2023. https://www.cnbc.com/2021/05/25/tesla-ditching-radar-for-autopilot-in-model-3-model-y.html

[5] Templeton, B. (12 December 2022). Tesla May Be Adding A Radar, Which Is The 'Crutch' They Definitely Need. Forbes. Last accessed: 30 April 2023. https://www.forbes.com/sites/bradtempleton/2022/12/12/tesla-may-be-adding-a-radar-which-is-the-crutch-they-definitely-need/?sh=eef16db73e7f

[6] Shuttleworth, J. (1 July 2019). SAE Standards News: J3016 automated-driving graphic update. SAE International. Last accessed: 30 April 2023. https://www.sae.org/news/2019/01/sae-updates-j3016-automated-driving-graphic

[7] McFarland, M. (17 January 2023). Tesla-induced pileup involved driver-assist tech, government data reveals. CNN Business. Last accessed: 30 April 2023. https://edition.cnn.com/2023/01/17/business/tesla-8-car-crash-autopilot/index.html

[8] Stempel, J. (15 September 2022). Tesla is sued by drivers over alleged false Autopilot, Full Self-Driving claims. REUTERS. Last accessed: 30 April 2023. https://www.reuters.com/business/autos-transportation/tesla-is-sued-by-drivers-over-alleged-false-autopilot-full-self-driving-claims-2022-09-14/

[9] Kolodny, L. (5 August 2022). California DMV accuses Tesla of deceptive practices in marketing Autopilot and Full Self-Driving options. CNBC. Last accessed: 30 April 2023. https://www.cnbc.com/2022/08/05/california-dmv-says-tesla-fsd-autopilot-marketing-deceptive.html

[10] Mitchell, R. (31 August 2022). Bill targeting Tesla's 'self-driving' claims passes California Legislature. Los Angeles Times. Last accessed: 30 April 2023. https://www.latimes.com/business/story/2022-08-31/newly-passed-bill-could-force-tesla-to-scrap-the-name-full-self-driving-in-california

[11] Gonzalez, L. (13 September 2022). An act to add Section 24011.5 to the Vehicle Code, relating to vehicles (Senate Bill No. 1398, CHAPTER 308). Last accessed: 30 April 2023. https://leginfo.legislature.ca.gov/faces/billTextClient.xhtml?bill_id=202120220SB1398

[12] National Assembly Secretariat. (30 March 2023a). National Assembly Minutes No. 2. In *The 404th National Assembly (Extraordinary Session)*. Last accessed: 30 April 2023. https://likms.assembly.go.kr/bill/billDetail.do?billId=PRC_Z2P3T0G2Z2M3V1Q9Y4N5J1V4O6X8L2

[13] Blanco, S. (8 March 2021). Autonomous Delivery Robots Are Now 'Pedestrians' in Pennsylvania. CAR AND DRIVER. Last accessed: 30 April 2023. https://www.caranddriver.com/news/a35756202/autonomous-delivery-robots-pedestrians-law/

[14] Said, C., Evangelista, B. (6 December 2018). San Francisco to robots: Don't crowd our sidewalks. San Francisco Chronicle. Last accessed: 30 April 2023. https://www.sfchronicle.com/news/article/San-Francisco-to-robots-Don-t-crowd-our-12411062.php

[15] Chairperson of the National Assembly Committee on Public Administration and Security. (28 March 2023). Partial Amendment Bill on Road Traffic Act (Alternative, Bill No. 20924). Last accessed: 30 April 2023. https://likms.assembly.go.kr/bill/billDetail.do?billId=PRC_Z2P3T0G2Z2M3V1Q9Y4N5J1V4O6X8L2

[16] Office For Government Policy Coordination. (19 July 2022). Explain "Unmanned delivery robots, which are still subject to absurd regulations that make it illegal to operate them without human companionship" (19 July, Maeil Business Newspaper). Last accessed: 30 April 2023. https://www.opm.go.kr/opm/news/press2.do?mode=view&articleNo=150601

[17] National Assembly Secretariat. (21 February 2023b). Committee on Public Administration and Security Minutes (Deliberation Subcommittee on Legislation No. 2). In *The 403rd National Assembly (Extraordinary Session)*. Last accessed: 30 April 2023. http://likms.assembly.go.kr/bill/billDetail.do?billId=PRC_K2X2Q0Y8S2I6H1G1V3N1U0H0S3X5L9&ageFrom=21&ageTo=21

[18] National Assembly Secretariat. (27 April 2023c). National Assembly Minutes. In: *The 405th National Assembly Minutes (Extraordinary Session)* (No. 5). Last accessed: 30 April 2023. http://likms.assembly.go.kr/bill/billDetail.do?billId=PRC_X2H2N0A8K1I2V0W9N2O2S1Z9F2B0C5&ageFrom=21&ageTo=21

[19] Park, K. C. (November 2022). Report on the review of the Partial Amendment Bill on Road Traffic Act. Last accessed: 30 April 2023. http://likms.assembly.go.kr/bill/billDetail.do?billId=PRC_K2X2Q0Y8S2I6H1G1V3N1U0H0S3X5L9&ageFrom=21&ageTo=21

[20] National Assembly Secretariat. (21 March 2023d). Legislation and Judiciary Committee Minutes No. 1. In *The 404th National Assembly (Extraordinary Session)*. Last accessed: 30 April 2023. https://likms.assembly.go.kr/bill/billDetail.do?billId=PRC_Z2P3T0G2Z2M3V1Q9Y4N5J1V4O6X8L2

[21] Lee, N. Y. (22 December 2021). Delivery robots stymied by regulation...risking missing out on golden time [Spectrum by Na Young Lee]. dailian. Last accessed: 30 April 2023. https://www.dailian.co.kr/news/view/1065701/?sc=Naver

[22] Namgoong, M. K. (20 February 2022). World's first D2D delivery robot drives... but regulations hold it back. edaily. Last accessed: 30 April 2023. https://www.edaily.co.kr/news/read?newsId=01580966632232488&mediaCodeNo=257

[23] Choi, A. R. (2 December 2022). After self-driving cars, the next step is 'robots'... securing data within the golden time is the key. koit. Last accessed: 30 April 2023. https://www.koit.co.kr/news/articleView.html?idxno=106326

[24] Yoon, H. S. (23 April 2021). Self-driving robots are on the streets~. Korea Policy Briefing. Last accessed: 30 April 2023. https://www.korea.kr/news/reporterView.do?newsId=148886260

[25] Kim, C. H. (18 October 2022). The delivery robot encountered a dog on the street. How would it react. Asia Economic. https://www.asiae.co.kr/article/2022101809422224347

[26] Park, J. W. (2019). DistilKoBERT: Distillation of KoBERT. *GitHub repository*. Last accessed: 4 May 2023. https://github.com/monologg/DistilKoBERT

[27] Lee, J. B. (2021). KcELECTRA: Korean comments ELECTRA. *GitHub repository*. Last accessed: 4 May 2023. https://github.com/Beomi/KcELECTRA

[28] Park, J. W., Kim, D. G. (2021). KoBigBird: Pretrained BigBird Model for Korean. *Zendo*. https://doi.org/10.5281/zenodo.5654154 Last accessed: 4 May 2023. https://github.com/monologg/KoBigBird

[29] Grootendorst., M. (2022b). I have only a few topics, how do I increase them?. *GitHub repository*. Last accessed: 4 May 2023. https://maartengr.github.io/BERTopic/faq.html#i-have-only-a-few-topics-how-do-i-increase-them

[30] Grootendorst, M. (2020). KeyBERT: Minimal keyword extraction with BERT. *Zenodo*. https://doi.org/10.5281/zenodo.4461265 Last accessed: 4 May 2023. https://github.com/MaartenGr/KeyBERT

[31] Lee, Y. W., & Yoo, Y. H. (2013). mekab-ko. Bitbucket. Last accessed: 4 May 2023. https://bitbucket.org/eunjeon/mecab-ko/src

[32] National Institute of Korean Language. (2020). ModuCorPus. Last accessed: 10 May 2023. https://corpus.korean.go.kr/?lang=en

[33] Kudo, T. (2006). MeCab. *SOURCEFORGE*. Last accessed: 12 May 2023. https://sourceforge.net/projects/mecab/

[34] Kim, H. J. (2019). SOYNLP: Python package for Korean natural language processing. *GitHub repository*. Last accessed: 13 July 2023. https://github.com/lovit/soynlp

[35] Ryu, W. H. (2014). twitter-korean-text. *GitHub repository*. Last accessed: 15 May 2023. https://github.com/twitter/twitter-korean-text/

[36] Shin, J. S., Park, J. H., & Lee, G. H. (2016). komoran. *GitHub repository*. Last accessed: 15 May 2023. https://github.com/shineware/KOMORAN

[37] Korea Legislation Research Institute. (2019). Product Liability Act. Last accessed: 1 June 2023. https://elaw.klri.re.kr/kor_service/lawView.do?hseq=43265&lang=ENG